\documentclass[aps,prc,showpacs,twocolumn,superscriptaddress]{revtex4}
\pdfoutput=1
\usepackage[usenames,dvipsnames]{color}
\usepackage{graphicx}
\usepackage{ulem}
\begin{document}

\title{Gas dynamics in high-luminosity polarized $^3$He targets using diffusion and convection}
\author{P.A.M. Dolph}
\affiliation{University of Virginia, Charlottesville, VA 22903}
\author{J. Singh}
\affiliation{University of Virginia, Charlottesville, VA 22903}
\affiliation{Argonne National Lab, Argonne, IL 60439}
\author{T.~Averett}
\affiliation{College of William and Mary, Williamsburg, VA 23187}
\author{A.~Kelleher}
\affiliation{College of William and Mary, Williamsburg, VA 23187}
\affiliation{Massachusetts Institute of Technology, Cambridge, MA 02139}
\author{K.E. Mooney}
\affiliation{University of Virginia, Charlottesville, VA 22903}
\author{V. Nelyubin}
\affiliation{University of Virginia, Charlottesville, VA 22903}
\author{W.A. Tobias}
\affiliation{University of Virginia, Charlottesville, VA 22903}
\author{B. Wojtsekhowski}
\affiliation{Thomas Jefferson National Accelerator Facility,
  Newport News, VA 23606} 
\author{G.D. Cates}
\thanks{Corresponding author: cates@virginia.edu} 
\affiliation{University of Virginia, Charlottesville, VA 22903}

\date{\today}

\begin{abstract}
The dynamics of the movement of gas is discussed for two-chambered polarized $^3$He target cells of the sort that have been used successfully for many electron scattering experiments.  A detailed analysis is presented  showing that diffusion is a limiting factor in target performance, particularly as these targets are run at increasingly high luminosities.  Measurements are presented on a new prototype polarized $^3$He target cell in which the movement of gas is due  largely to convection instead of diffusion.  NMR tagging techniques have been used to visualize the gas flow, showing velocities along a cylindrically-shaped target of between $\rm 5-80\,cm/min$.  The new target design addresses one of the principle obstacles to running polarized $^3$He targets at substantially higher luminosities while simultaneously providing new  flexibility in target geometry.
\end{abstract}

\maketitle

\section{Introduction}
Nuclear-polarized $^3$He has proven to be useful in a number of different areas of research.  In electron scattering, polarized $^3$He provides a means for studying spin-dependent interactions involving neutrons.  This is because, to first approximation, a $^3$He nucleus is comprised of two protons whose spins are paired, and a single neutron that accounts for most of the nuclear spin~\cite{fri90}.   An important early example of the use of polarized $^3$He in electron scattering came during an experiment to measure the internal spin structure of the neutron at the Stanford Linear Accelerator Center (SLAC), E142~\cite{PhysRevLett.71.959}.  Polarized $^3$He has also been used to measure the electric form factor of the neutron $G_E^n$, including a recent experiment at Jefferson Laboratory (JLab) in Newport News, VA~\cite{PhysRevLett.105.262302}.  Important applications of polarized $^3$He have also included its use as a neutron polarizer~\cite{Coulter1990463}, and, together with polarized $^{129}$Xe, as a source of signal for magnetic resonance imaging~\cite{Nature94,MRimaging}.

There are two predominant techniques by which high levels of nuclear polarization are produced in  $^3$He.  In one technique, often known as metastability exchange optical pumping (MEOP), metastable states of $^3$He are optically pumped directly, and subsequently transfer their polarization to other ground-state $^3$He nuclei during metastability-exchange collisions~\cite{colegrove1963,nacher1985}.  In a second technique, known as spin-exchange optical pumping (SEOP), a vapor of alkali-metal atoms is optically pumped and subsequently transfers its polarization to $^3$He nuclei via a hyperfine interaction during  spin-exchange collisions~\cite{bouchiat1960,bhaskar1982,PhysRevC.36.2244}.  An important  difference between the two techniques is that MEOP is performed at pressures that are quite low, around a few Torr, whereas SEOP is often done at pressures as high as roughly ten atmospheres.  When a high-density is required, a target based on MEOP inevitably involves a compressor of some sort.  In contrast, high-density targets based on SEOP typically utilize a sealed glass cell with no moving parts and hence have an advantage from the perspective of simplicity.  In considering relative merits, however, one also needs to consider the speed with which the gas is polarized.  Here targets based on MEOP have done quite well, with a recent target at the Mainz Microtron reporting a polarization rate of 2 bar-liters per hour\cite{Krimmer200918}.  In short, both techniques for polarizing $^3$He have been quite successful and have complimentary advantages.

\begin{figure}[ht]
\begin{center}
\hbox{\hskip -0.11truein \includegraphics[width = 9.4cm]{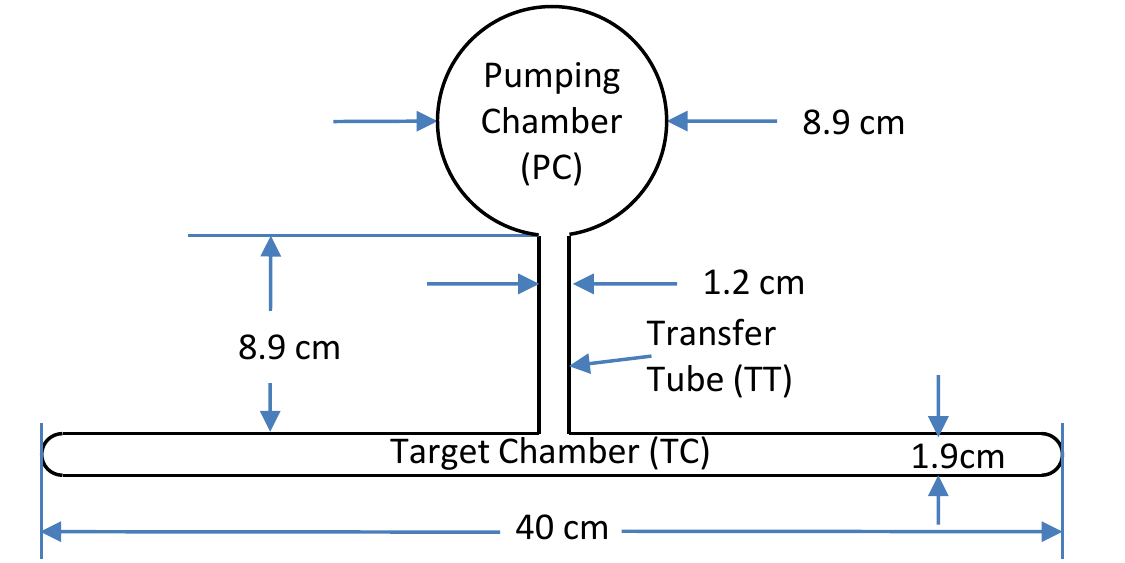}}
\caption{Shown is the geometry of a two-chambered glass cell used for polarized $^3$He targets.  The dimensions shown are typical of those used in a recent JLab experiment to measure the electric form factor of the neutron
\cite{PhysRevLett.105.262302}.}
\label{fig:genstyle}
\end{center}
\end{figure}

For electron-scattering experiments in which $^3$He is polarized using SEOP, the targets typically 
utilize a sealed glass cell with two distinct chambers: a ``pumping chamber" in which the gas is polarized, and a ``target chamber" through which the electron beam passes (see Fig.~\ref{fig:genstyle}).  This design ensures that ionization due to the electron beam does not adversely affect the optical pumping process, as well as providing in the pumping chamber a geometry that lends itself well to illumination with lasers.   The two chambers are connected by a ``transfer tube", and gas that is polarized in the pumping chamber migrates into the target chamber largely by diffusion.  

Several conditions need to be met in targets with designs such as that shown in Fig.~1 in order to maintain high polarization.  First, the rate at which the $^3$He is polarized must be relatively fast compared to the rate at which the $^3$He is depolarized.  While this is true in all work involving SEOP, it is of particular significance when considering depolarization of the $^3$He due to an electron beam.  There are also issues having to do with polarization dynamics that are unique to the two-chamber design.   As has been pointed out by Chupp {\it et al.}\cite{PhysRevC.45.915}, it is important that diffusion between the two chambers is rapid compared to other $^3$He-related time constants in the system.  As we will show in detail, a matter of special importance is maximizing the ratio of the diffusion rate to the relaxation rate that is specific to the target chamber.  Put differently, it is critical that the polarized gas in the target chamber is replenished much faster than it is depleted through depolarization mechanisms such as, for example, ionization from the electron beam.  A failure to replenish the polarized gas quickly enough will result in a lower polarization in the target chamber than is the case in the pumping chamber, a condition we refer to herein as a {\it polarization gradient}.   

Until relatively recently, polarization gradients in two-chambered cells have been at most a few percent relative.   Advances in SEOP, however, have made it possible to run polarized $^3$He targets at increasingly high luminosities.  During SLAC E142, where luminosities were in the range of $\rm 0.42-1.70\times10^{35}\,cm^{-2}s^{-1}$, the polarization gradients were on the order of 1\%.  During recent experiments at JLab, however,  with luminosities in the range of $\rm 0.6-1.0\times10^{36}\,cm^{-2}s^{-1}$, polarization gradients were as high as nearly 9\% (relative).  For future experiments with luminosities in the range of  $\rm 10^{37}\,cm^{-2}s^{-1}$, polarization gradients well in excess of 10\% are likely if no changes are made to the basic target-cell design.  Polarization gradients can also be difficult to quantify accurately, an issue that can lead to uncertainties in polarimetry.

The fact that polarized $^3$He targets are being run at increasingly high luminosities is due largely to advances  in SEOP.  One example is the use of hybrid mixtures of alkali metals (typically potassium and rubidium) instead of a single alkali metal (typically rubidium)\cite{happer2001,babcock2003}.  This technique, often known as alkali-hybrid SEOP, greatly improves the efficiency with which the $^3$He is polarized. Another important advance has followed from the availability of commercial spectrally-narrowed high-power diode-laser arrays.  These new lasers result in significantly higher optical pumping rates for a given amount of light.  Collectively, these advances have made it possible to significantly increase the rate at which the $^3$He is polarized.  It has thus become possible to achieve higher polarizations even when using higher beam currents and thicker targets.  Higher currents, however, make it necessary to replenish the polarized gas in the target chamber more quickly.  If the advances in SEOP are to be fully exploited, it is essential that the designs of polarized $^3$He targets evolve.

We report here on the successful implementation of a new design for polarized $^3$He target cells based on SEOP.  The design incorporates the ability to circulate the gas between the pumping chamber and the target chamber using convection instead of diffusion, an idea discussed by Wojtsekhowski in Ref.~\cite{woj02} in anticipation of both advances in SEOP as well as the need for higher luminosities.  The convection is achieved by maintaining a temperature differential between different parts of the target cell, and does not involve pumps or other moving parts.  We have shown that the velocity of the gas moving through the target chamber can be varied between 5--80$\,\rm cm/min$ in a simple controllable manner.  The advent of a means to circulate a polarized noble gas in a sealed vessel without the use of pumps has great potential for high-luminosity polarized $^3$He targets.  The simplicity of the approach has advantages from the perspective of reliability, and fast transport of gas between the two chambers makes it possible to greatly increase the electron beam current without causing a polarization gradient.  Convection-based target cells also open the possibility of physically separating the pumping chamber and the target chamber by much larger distances than was previously possible, something that offers several practical advantages.   

\section{Polarization Dynamics and Gradients in  Two-Chambered Cells}

\subsection{The Single-Chambered Cell}

Before considering the formalism for target cells with two chambers, we begin by considering the simpler example of a single-chambered cell, where the equation describing the time evolution of the polarization is given by:
\begin{equation}
\dot{P}_{\rm He} = \gamma_\mathrm{se} P_\mathrm{A} - (\gamma_\mathrm{se}+\Gamma) P_{\rm He},
\label{eq:singlecell}
\end{equation}
where $P_{\rm He}$ is the $^3$He polarization, $P_\mathrm{A}$ is the polarization of the alkali vapor, $\gamma_\mathrm{se}$ is the rate at which the $^3$He is polarized due to spin exchange, and $\Gamma$ is the spin-relaxation rate of the $^3$He due to all other processes.  The solution to Eqn.\ref{eq:singlecell} is given by 
\begin{equation}
P_{\rm He}(t) = P_\mathrm{0}e^{-(\gamma_\mathrm{se} + \Gamma)t} + P_\mathrm{A}{{\gamma_\mathrm{se}}\over{\gamma_\mathrm{se} + \Gamma}}(1-e^{-(\gamma_\mathrm{se}+\Gamma)t})
\label{eqn:psingle}
\end{equation}
where $P_\mathrm{0}$ is the $^3$He polarization at $t=0$.  It has been shown by Babcock {\it et al.} that one of the components of $\Gamma$ is a relaxation rate that empirically appears to be proportional to $\gamma_\mathrm{se}$~\cite{babcock2006}.  One can accordingly write that $\Gamma = \Gamma^{\prime} + \gamma_\mathrm{se}X$, where $X$ is a proportionality constant.  We can thus write the saturation polarization associated with Eqn.~\ref{eqn:psingle} as
\begin{equation}
P_{\rm He}(t=\infty) = P_\mathrm{A}{{\gamma_\mathrm{se}}\over{\gamma_\mathrm{se}(1+X) + \Gamma^{\prime}}} \ \ ,
\label{eqn:psingle_w_x}
\end{equation}
where we note that the denominator of Eqn.~\ref{eqn:psingle_w_x} is also the rate that characterizes the buildup of polarization in \ref{eqn:psingle}, rewritten in terms of $\Gamma^{\prime}$ and $X$.  The existence of a relaxation mechanism proportional to $\gamma_\mathrm{se}$ is unfortunate, as it implies that, even with a 100\% polarized alkali vapor, in the limit that $\gamma_\mathrm{se}\rightarrow \infty$, $P_{\rm He} \rightarrow 1/(1+X)$.   While not well understood, the existence of this additional relaxation is now well established, and is important for understanding the overall polarization achieved.

It is straightforward to project the performance of a target at an arbitrary range of beam currents if we know how it performed at a particular beam current $I_\mathrm{0}$ (true even if $I_\mathrm{0} = 0$).  Let us assume that we have data describing polarization as a function of time while polarizing a target from an initial polarization of zero.  A plot of data of this sort is something we refer to herein as a ``spin-up curve", and can be fit to yield the ``spin-up rate" that appears in the denominator of Eqn.~\ref{eqn:psingle_w_x} (as well as the exponent in Eqn.~\ref{eqn:psingle}, $\gamma_\mathrm{se} + \Gamma$).  In keeping with the notation introduced above with respect to Eqn.~\ref{eqn:psingle_w_x}, we define a quantity $\gamma_\mathrm{su}^0$  that characterizes the spin-up rate at a particular current $I_0$:
\begin{equation}
\gamma_\mathrm{su}^0 \equiv \gamma_\mathrm{se}(1+X) + \Gamma^{\prime}(I_\mathrm{0})\ \ .
\label{eq:gamma_su_def}
\end{equation}
Let us further define $P_{\rm He}^{\infty}(I_\mathrm{0})$ as the equilibrium polarization associated with that spin-up.  Using Eqns.~\ref{eqn:psingle_w_x} and \ref{eq:gamma_su_def}, we find
\begin{equation}
P_{\rm He}^{\infty}(I) = {{P_{\rm He}^{\infty}(I_\mathrm{0})\,\gamma_\mathrm{su}^0}\over{\gamma_\mathrm{su}^0 - \Gamma^{\prime}(I_\mathrm{0}) + \Gamma^{\prime}(I)}}\ \ .
\label{eq:model}
\end{equation}
The quantities $P_{\rm He}^{\infty}(I_\mathrm{0})$ and $\gamma_\mathrm{su}^0$ can be determined by fitting data from a spin up, and  the quantity $\Gamma^{\prime}(I) = \Gamma_\mathrm{wall}^\mathrm{cold} + \Gamma_\mathrm{dipole} + \Gamma_\mathrm{beam}$, 
where the three terms are spin-relaxation rates due to wall collisions (at room temperature), dipole interactions due to  $^3$He-$^3$He collisions, and  ionization of the electron beam respectively.  The sum of the first two terms is essentially the target's room-temperature spin-relaxation rate in the absence of beam, a quantity that is quite easy to measure.  We note, however, that $\Gamma_\mathrm{dipole}$ will be slightly different under operating conditions than it is at room temperature both because of heating the pumping chamber and the related differences in densities.  This correction is easily calculated using the calculation of Newbury {\it et al.} from Ref.~\cite{new93}.  The contribution $\Gamma_\mathrm{beam}$ is also easily computed~\cite{PhysRevA.37.3270,PhysRevA.38.4481}.  We note that calculations of  $\Gamma_\mathrm{beam}$ have shown good agreement with experiment at a level of roughly 10\% or better~\cite{Coulter198929}.   We can thus use Eqn.~\ref{eq:model}  to project the performance of a particular target at an arbitrary beam current.

It is instructive to investigate how different targets that have been used during past experiments would fare at significantly increased current, for example, at $100\,\mu\rm A$.  Here we ignore the effect of polarization gradients.  The first time liter-type quantities of polarized $^3$He were used in a target was the aforementioned experiment at SLAC, E142~\cite{PhysRevLett.71.959}.  
In the presence of $3.3\,\mu\rm A$ of electron beam, the $^3$He polarization averaged about 33\%, with values for $(\gamma_\mathrm{su}^0)^{-1}$ of about 15-20 hours.  If this same target were instead exposed to $100\,\mu\rm A$ of electron beam, Eqn.~\ref{eq:model} suggests that the $^3$He polarization would drop to just over 10\%.
In contrast,  during the more recent experiment at JLab that measured $G_E^n$~\cite{PhysRevLett.105.262302}, the cell-averaged polarizations were around 50\% with  $8\,\mu\rm A$ of beam and $(\gamma_\mathrm{su}^0)^{-1}$ was in the range of $5-6\,\rm hrs$.  Here Eqn.~\ref{eq:model} suggests that at   $100\,\mu\rm A$ the resulting $^3$He polarization would be around 37--38\%.
The improved projected performance is due both to the shorter values for $(\gamma_\mathrm{su}^0)^{-1}$, as well as the fact that the target cells were significantly larger, making them more resistant to depolarization from the electron beam.  We note, however, that even though the cell-averaged polarization would be fairly reasonable, the polarization that one would have in the target chamber would be much lower because of polarization gradients.

\subsection{Time Evolution in a Double-Chambered Cell}
\label{section:DCC}

For a full description of a double-chambered cell, the polarization build up must be described by the coupled differential equations first described in Ref.~\cite{PhysRevC.45.915}:
\begin{eqnarray}
\dot{P}_\mathrm{pc} &=& \gamma_\mathrm{se}(P_\mathrm{A}-P_\mathrm{pc})-\Gamma_\mathrm{pc}P_\mathrm{pc}-d_\mathrm{pc}(P_\mathrm{pc}-P_\mathrm{tc}) \label{eqn:ppcdot1} \\
\dot{P}_\mathrm{tc} &=& -\Gamma_\mathrm{tc}P_\mathrm{tc}+d_\mathrm{tc}(P_\mathrm{pc}-P_\mathrm{tc})
\label{eqn:ptcdot1}
\end{eqnarray}
where $P_\mathrm{pc}$ ($P_\mathrm{tc}$) is the $^3$He polarization in the pumping (target) chamber, $\gamma_\mathrm{se}$ is the spin-exchange rate in the pumping chamber, and $\Gamma_\mathrm{pc}$ and  $\Gamma_\mathrm{tc}$ are the $^3$He spin-relaxation rates due to all other processes in the pumping and target chambers respectively.  The transfer rate $d_\mathrm{tc}$ ($d_\mathrm{pc}$) is the probability per unit time per nucleus that a nucleus will exit the target (pumping) chamber and enter the pumping (target) chamber.  We note that we do not include the transfer tube as a separate volume in this analysis.  The transfer rates are related by 
\begin{equation}
f_\mathrm{pc}d_\mathrm{pc} = f_\mathrm{tc}d_\mathrm{tc}
\label{eqn:dRelation}
\end{equation}
where $f_\mathrm{pc}$($f_\mathrm{tc}$) is the fraction of atoms in the pumping (target) chamber, and $f_\mathrm{pc}+f_\mathrm{tc} = 1$.
For the dynamic studies reported in Ref.~\cite{PhysRevC.45.915}, the authors were able to neglect terms involving $\gamma_\mathrm{se}$ and $\Gamma$ relative to terms involving $d_\mathrm{pc}$ and $d_\mathrm{tc}$.  For the discussion here, however, we must retain these terms, requiring an analysis essentially identical to that considered by Jones {\it et al.}~\cite{jon93} and later by Kominis~\cite{kominis}.  We refer the reader to those two references for details.  We 
find that the solutions to Eqns.~\ref{eqn:ppcdot1} and \ref{eqn:ptcdot1} are given by 
\begin{equation}
P_\mathrm{pc}(t) = C_\mathrm{pc}e^{-\Gamma_\mathrm{f}t} + (P_\mathrm{pc}^0-P_\mathrm{pc}^\infty - C_\mathrm{pc})e^{-\Gamma_\mathrm{s}t} + P_\mathrm{pc}^\infty
\label{eqn:ppct}
\end{equation}
and
\begin{equation}
P_\mathrm{tc}(t) = C_\mathrm{tc}e^{-\Gamma_\mathrm{f}t} + (P_\mathrm{tc}^0-P_\mathrm{tc}^\infty - C_\mathrm{tc})e^{-\Gamma_\mathrm{s}t} + P_\mathrm{tc}^\infty
\label{eqn:ptct}
\end{equation}
where $P_\mathrm{pc}^0$ and $P_\mathrm{tc}^0$ are the initial polarizations in the  pumping and target chambers respectively, 
\begin{equation}
P_\mathrm{pc}^\infty = {{\gamma_\mathrm{se}\,f_\mathrm{pc}\,P_\mathrm{A}}\over{\gamma_\mathrm{se}\,f_\mathrm{pc} + \Gamma_\mathrm{pc}\,f_\mathrm{pc} + \Gamma_\mathrm{tc}\,f_\mathrm{tc}(1+{{\Gamma_\mathrm{tc}}\over{d_\mathrm{tc}}})^{-1}}}\ \ ,
\label{eq:pcsatpol}
\end{equation}
and 
\begin{equation}
P_\mathrm{tc}^\infty = P_\mathrm{pc}^\infty\,{1\over{1+{{\Gamma_\mathrm{tc}}\over{d_\mathrm{tc}}}}} \ \ .
\label{eq:polgrad}
\end{equation}
We have chosen to write Eqn.~\ref{eq:pcsatpol} in the form shown to emphasize that
\begin{equation}
\lim_{(\Gamma_\mathrm{tc}/d_\mathrm{tc}) \to 0}{P_\mathrm{tc}^\infty}= \lim_{(\Gamma_\mathrm{tc}/d_\mathrm{tc}) \to 0}{P_\mathrm{pc}^\infty}= {{P_\mathrm{A}\,\langle\gamma_\mathrm{se}\rangle}\over{\langle\gamma_\mathrm{se}\rangle +\langle\Gamma\rangle}}
\label{eq:limit}
\end{equation}
where $\langle \gamma_\mathrm{se} \rangle$ is the spin-exchange rate averaged throughout the double-chambered cell ($\langle \gamma_\mathrm{se} \rangle= f_\mathrm{pc}\gamma_\mathrm{se}$,  since the spin-exchange rate is $\gamma_\mathrm{se}$ in the pumping chamber and is zero in the target chamber), and  $\langle \Gamma \rangle = f_\mathrm{pc}\Gamma_\mathrm{pc} + f_\mathrm{tc}\Gamma_\mathrm{tc}$ is the spin-relaxation rate averaged throughout the cell.  Eqn.~\ref{eq:limit} has the same form as the saturation polarization of Eqn.~\ref{eqn:psingle}, as we would expect in the limit of infinitely fast transfer.   

The coefficients $C_\mathrm{pc}$ and $C_\mathrm{tc}$ are given by
\begin{equation}
C_\mathrm{pc} = \frac{\Gamma_\mathrm{s}(P_\mathrm{pc}^\infty-P_\mathrm{pc}^0)-aP_\mathrm{pc}^0-bP_\mathrm{tc}^0- \gamma_\mathrm{se}P_\mathrm{A}}{\Gamma_\mathrm{f}-\Gamma_\mathrm{s}}\ \ .
\end{equation}
and
\begin{equation}
C_\mathrm{tc} = \frac{\Gamma_\mathrm{s}(P_\mathrm{tc}^\infty-P_\mathrm{tc}^0)-cP_\mathrm{pc}^0-dP_\mathrm{tc}^0}{\Gamma_\mathrm{f}-\Gamma_\mathrm{s}}\ \ ,
\end{equation}
where $a = -(\gamma_\mathrm{se} + \Gamma_\mathrm{pc} + d_\mathrm{pc})$,
$b=d_\mathrm{pc}$,
$c= d_\mathrm{tc} $ and
$d = -(\Gamma_\mathrm{tc} + d_\mathrm{tc})$.
We note that in the fast-transfer limit, the quantities $C_\mathrm{pc}$ and $C_\mathrm{tc}$ are given by
\begin{equation}
C_\mathrm{pc} = f_\mathrm{tc}(P_\mathrm{pc}^0 - P_\mathrm{tc}^0) \ ,
\end{equation}
and
\begin{equation}
C_\mathrm{tc} = f_\mathrm{pc}(P_\mathrm{tc}^0 - P_\mathrm{pc}^0) \ .
\end{equation}

The constants $\Gamma_\mathrm{s}$ and $\Gamma_\mathrm{f}$ represent slow and fast rates respectively that govern the time evolution of the polarization.  It is useful to write $\Gamma_\mathrm{s}$ in the form
\begin{equation}
\Gamma_\mathrm{s} \ =  \langle\gamma_\mathrm{se}\rangle + \langle\Gamma\rangle -  \delta\Gamma \ 
\label{eq:gamma_slow_explained}
\end{equation}
where the quantity $\delta\Gamma$ is generally small and goes to zero in the limit of infinitely fast transfer between the two chambers.  We can see that the limiting form of $\Gamma_\mathrm{s}$ has the same form as the time constant that appears in Eqn.~\ref{eqn:psingle} ($\gamma_\mathrm{se}+\Gamma$).   It can be shown that
\begin{equation}
\delta\Gamma = \frac{d_\mathrm{pc}+d_\mathrm{tc}}{2}\left[\sqrt{1-2u\delta f+u^2}-1+u\delta f\right] 
\end{equation}
where $\delta f = f_\mathrm{pc}-f_\mathrm{tc}$ and
\begin{equation}
u = \frac{\gamma_\mathrm{se} + \Gamma_\mathrm{pc} - \Gamma_\mathrm{tc}}{d_\mathrm{pc}+d_\mathrm{tc}} \ \ .
\end{equation}
For most of the situations we would normally consider, the quantity $u$ is fairly small.  This is due to two things.  First, the spin-exchange rate $\gamma_\mathrm{se}$ is typically slow compared to the sum of the two transfer rates $d_\mathrm{pc}$ and $d_\mathrm{tc}$, and second, both  $\Gamma_\mathrm{pc}$ and $\Gamma_\mathrm{tc}$ must be relatively small compared to $\gamma_\mathrm{se}$, or the polarization of the target would not be high.   
It is thus reasonable to expand $\delta\Gamma$ in a Taylor series in $u$:
\begin{equation}
\delta\Gamma \approx f_\mathrm{pc}f_\mathrm{tc}(d_\mathrm{pc}+d_\mathrm{tc})u^2 + \rm higher\ order\ terms.
\end{equation}
Lastly, we consider $\Gamma_\mathrm{f}$ which can be written as
\begin{equation}
\Gamma_\mathrm{f} = (d_\mathrm{pc}+d_\mathrm{tc}) + (\gamma_\mathrm{se}-\langle \gamma_\mathrm{se}\rangle) + (\Gamma_\mathrm{pc}+\Gamma_\mathrm{tc}-\langle \Gamma \rangle) + \delta\Gamma\ \ .
\end{equation}
In the fast-transfer limit, $\Gamma_\mathrm{f} \rightarrow \infty$; under these conditions, Eqns.~\ref{eqn:ppct}  and \ref{eqn:ptct} reduce to the form of Eqn.~\ref{eqn:psingle}.

Data illustrating the time evolution of polarization (what we referred to earlier as a spin-up) are shown in Fig.~\ref{fig:brady} for both the pumping and target chambers of a double-chambered cell we refer to herein as ``Brady".  The polarization was measured every three minutes using the NMR technique of adiabatic fast passage (AFP)~\cite{Abragam}.  We note that under normal operating conditions, NMR measurements would only be made once every few hours, in part because each measurement results in a small loss ($<1\%$) of polarization.  The frequent measurements shown in Fig.~\ref{fig:brady} strongly limit the saturation polarization because of accumulating losses.   Also shown in Fig.~\ref{fig:brady}, but obscured beneath the many data points, is a fit to the data using double-exponential functions of the form given in Eqns.~\ref{eqn:ppct}  and \ref{eqn:ptct}.  The fit clearly describes the data quite well.

Finally, we note that in the context of the types of cells that have been used in electron-scattering experiments (see Fig.~\ref{fig:genstyle}), the mechanism behind the transfer rates $d_\mathrm{tc}$ and $d_\mathrm{pc}$ is overwelmingly diffusion.

\subsection{Initial Polarization Evolution}

\begin{figure}
\begin{tabular}{l}
\includegraphics[width = 8.6 cm]{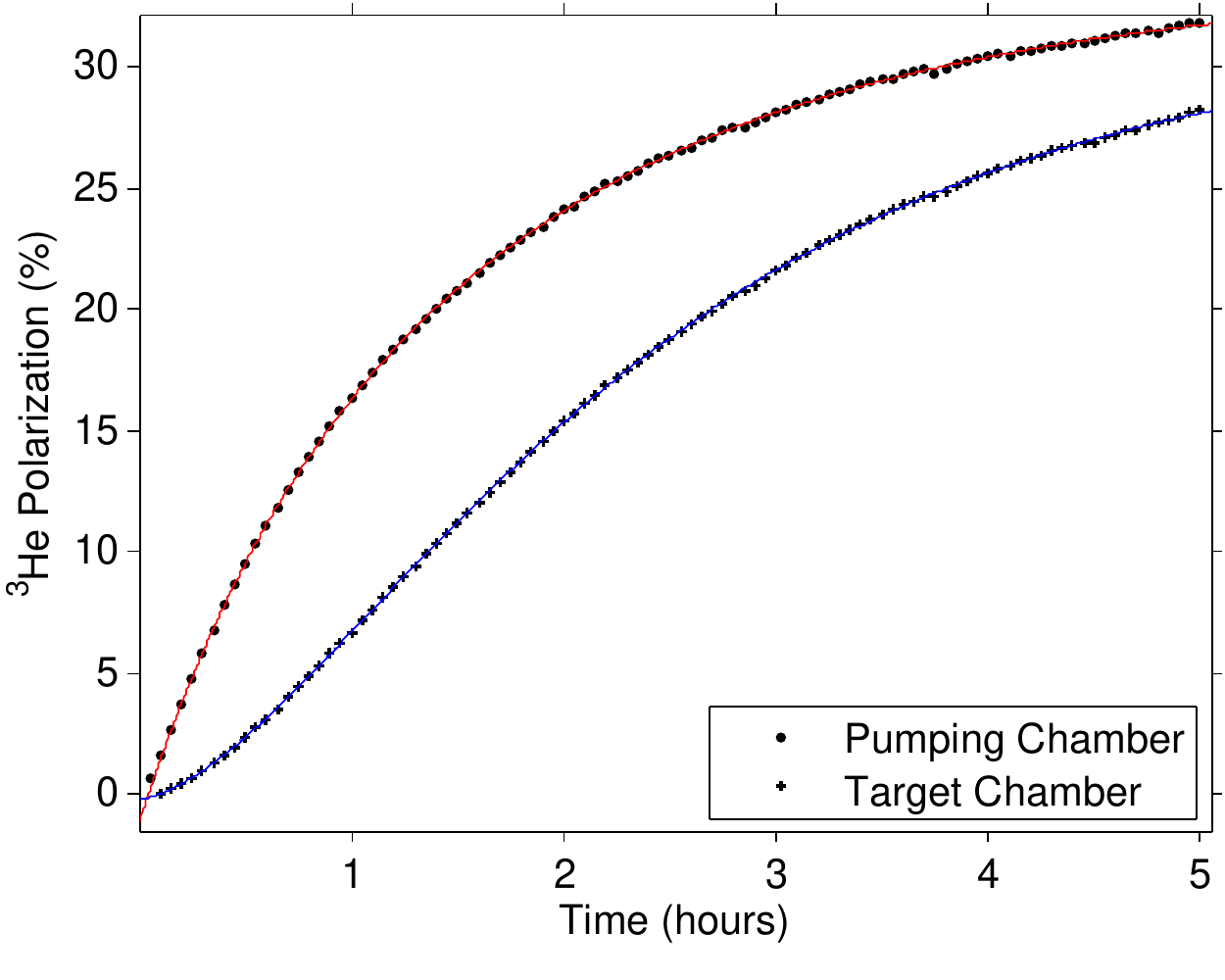}\\
\end{tabular}
\vskip -0.2truein
\caption{The $^3$He polarization is shown as a function of time for both the pumping chamber (upper curve) and target chamber (lower curve) of the target cell ``Brady".  In this case the lasers were turned on immediately before data taking to ensure an initial polarization of zero.  Also shown are fits to Eqns.~\ref{eqn:ppct} and \ref{eqn:ptct}.  We refer to curves of this sort as spin-up curves.  AFP measurements were made rapidly (every 3 minutes).}
\label{fig:brady}
\end{figure}

\begin{figure}[hb]
\vskip -0.15truein
\begin{center}
\includegraphics[width = 8.6 cm]{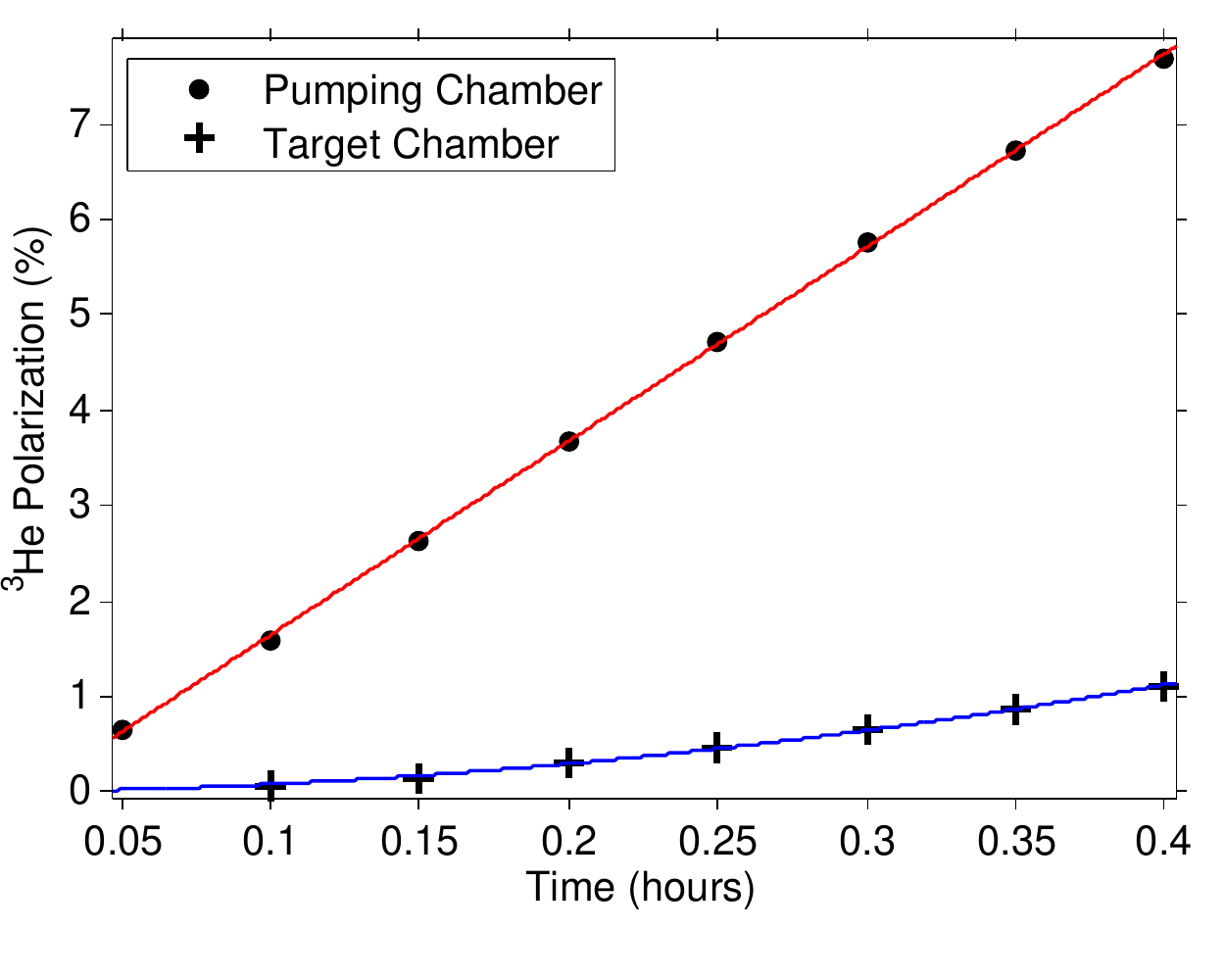}
\vskip -0.2truein
\caption{Early time behavior of the Brady spin-up shown in Fig.~\ref{fig:brady}.}
\label{fig:bradyEarly}
\end{center}
\end{figure}

Some of the parameters discussed earlier can be readily determined by studying spin-up curves of the sort shown in Fig.~\ref{fig:brady}.  To extract values for the transfer rates $d_\mathrm{pc}$ and  $d_\mathrm{tc}$, it is particularly valuable to examine the spin-up curves for the initial time period during which the polarization is growing.  For small values of the time $t$, it is readily apparent from Fig.~\ref{fig:brady} that the nature of the time evolution in the two chambers is quite different.  Under the assumption that the time $t < 1/\Gamma_\mathrm{f}$ (in this case, $1/\Gamma_\mathrm{f} \approx 0.75$hrs), we can expand Eqns.~\ref{eqn:ppct} and \ref{eqn:ptct} in a Taylor series.  To second order, for the case of  $P_\mathrm{pc}^0 = P_\mathrm{tc}^0 = 0$, this expansion simplifies to:
\begin{equation}
P_\mathrm{pc}(t) = \gamma_\mathrm{se}P_\mathrm{A}t - \frac{1}{2}\gamma_\mathrm{se}P_\mathrm{A}(\gamma_\mathrm{se}+\Gamma_\mathrm{pc}+d_\mathrm{pc})t^2  \label{eq:taylorpc}
\end{equation}
and 
\begin{equation}
P_\mathrm{tc}(t) = \frac{1}{2}\gamma_\mathrm{se}P_\mathrm{A}d_\mathrm{tc}t^2 \, . \label{eq:taylortc}
\end{equation}
In Fig.~\ref{fig:bradyEarly}, we show only the first 24 minutes of the data shown in Fig.~\ref{fig:brady}.  It can be seen that the initial shape of the spin-up curve appears to be linear in the pumping chamber and quadratic in the target chamber, in agreement with expectations from Eqns.~\ref{eq:taylorpc} and \ref{eq:taylortc}.  

To empirically determine $d_\mathrm{tc}$, we note first that the slope of the nearly-linear polarization buildup in the pumping chamber is equal to the product $P_\mathrm{A}\gamma_\mathrm{se}$.  With a fit to this slope, along with a fit to the coefficient characterizing the quadratic polarization buildup in the target chamber, we can extract a value for the  transfer rate $d_\mathrm{tc} = (0.72\pm 0.10)$hrs$^{-1}$.   As will be discussed in the next subsection, a value for $d_\mathrm{tc}$ can also be computed from first principles given the dimensions of the cell.  The comparison of empirically determined and calculated values for $d_\mathrm{tc}$ provides insight into our understanding of the diffusion processes taking place in our cells.

\subsection{Transfer Rates Under Diffusion}
Using gas kinetic theory, the dimensions of the target cell,  the density of $^3$He and other gases when the cell was filled, the operating temperatures of the pumping and target chambers and the assumption that the temperature gradient along the transfer tube is linear, it possible to compute $d_\mathrm{pc}$ and $d_\mathrm{tc}$ from first principles.  We begin similarly to the discussion in Ref.~\cite{PhysRevC.45.915} with an equation describing the net polarization flux $J_\mathrm{tt}$ through the transfer tube:
\begin{equation}
J_\mathrm{tt} = -n(z)\,D(z){{dP(z)}\over{dz}}
\label{eq:diff_diff_eq}
\end{equation}
where $n(z)$,  $D(z)$ and $P(z)$ are the $^3$He number density, self-diffusion constant and polarization respectively, all shown as a function of position $z$ along the transfer tube.    As discussed by Romalis~\cite{romulus} and later by Zheng~\cite{Zheng}, using data on the self-diffusion constant of $^4$He by Kestin {\it et al.}~\cite{JPCRD},  the self diffusion constant for $^3$He is well approximated by the expression
\begin{equation}
D(z) = D_\mathrm{0}\,\left({{T(z)}\over{T_\mathrm{0}}}\right)^{m-1}\left({{n_\mathrm{0}}\over{n(z)}}\right)
\label{eq:diff_constant}
\end{equation}
where $D_\mathrm{0}=\rm 2.789\,cm^2/s$, $T_\mathrm{0}=353.14\,$K, $m=1.705$ and $n_\mathrm{0}=\rm 0.7733\,amg$ (1 amg = $2.687\times 10^{19}$cm$^{-3}$). 
The assumption of a linear temperature gradient along the transfer tube between the pumping and target chambers, as was assumed in Refs.~\cite{romulus,Zheng}, allows us to express $T(z)$ and hence $n(z)$ explicitly.   With this assumption, and substituting Eqn.~\ref{eq:diff_constant} into Eqn.~\ref{eq:diff_diff_eq}, we can solve for $J_\mathrm{tt}$ to find
\begin{equation}
J_\mathrm{tt} = -(P_\mathrm{pc}-P_\mathrm{tc})D_\mathrm{0}\,      {{n_\mathrm{0}\,(2-m)(T_\mathrm{pc}-T_\mathrm{tc})}\over{L_\mathrm{tt}(T_\mathrm{0}^{m-1})(T_\mathrm{pc}^{2-m}-T_\mathrm{tc}^{2-m})}}\ .
\end{equation}
$J_\mathrm{tt}$ is the total rate of polarization transfer per unit area, whereas we want the rate per atom.  Multiplying by the transfer tube cross sectional area $A_\mathrm{tt}$, dividing by the number of particles in each chamber, and finally dividing by $(P_\mathrm{pc}-P_\mathrm{tc})$, we have
\begin{equation}
d_\mathrm{tc(pc)} = \frac{A_\mathrm{tt}\,D_\mathrm{0} }{V_\mathrm{tc(pc)}L_\mathrm{tt}} \,   {{n_\mathrm{0}\,(2-m)(T_\mathrm{pc}-T_\mathrm{tc})}\over{n_\mathrm{tc(pc)}(T_\mathrm{0}^{m-1})(T_\mathrm{pc}^{2-m}-T_\mathrm{tc}^{2-m})}}\ .
\label{eqn:dtc}
\end{equation}

Of specific interest here is the value for $d_\mathrm{tc}$ implied by Eqn.~\ref{eqn:dtc} for the target cell Brady.  For Brady, $A_\mathrm{tt} = 0.667$cm$^2$, $L_\mathrm{tt} = 9.07$cm, $V_\mathrm{tc} = 74.6\,$cm$^3$, and $n_\mathrm{tc} =\rm 11.5\, amg$.  Using temperatures that correspond to the tests illustrated in  Figs.~\ref{fig:brady} and \ref{fig:bradyEarly}, we find $d_\mathrm{tc} = 0.72\,$hrs$^{-1}$, a value that agrees fortuitously with the value found by fitting the polarization buildup curves. 

\subsection{Polarization Gradients}

An issue of considerable practical importance for polarized $^3$He targets is the polarization gradient between the pumping and target chambers.  Dividing both sides of Eqn.~\ref{eq:polgrad} by ${P_\mathrm{pc}^\infty}$, we find
\begin{equation}
\frac{P_\mathrm{tc}^\infty}{P_\mathrm{pc}^\infty} = \frac{1}{1+\Gamma_\mathrm{tc}/d_\mathrm{tc}}.
\label{eqn:ratio}
\end{equation}
It is also useful to define the quantity $\Delta$, the amount by which the polarization in the target chamber is lower than that of the pumping chamber.  Here we express this difference as a fraction of the pumping chamber polarization:
\begin{equation}
\Delta \equiv \frac{P_\mathrm{pc}^\infty - P_\mathrm{tc}^\infty}{P_\mathrm{pc}^\infty} = \frac{1}{1+d_\mathrm{tc}/\Gamma_\mathrm{tc}}
\label{eq:delta}
\end{equation}
We can see that $\Delta$ approaches 0 for cells in which $\Gamma_\mathrm{tc}$ is quite slow and $d_\mathrm{tc}$ is quite fast.  The fact that it is $\Gamma_\mathrm{tc}$ and not the cell-averaged relaxation rate $\langle \Gamma \rangle$ that appears in this expression is quite important.  When a target is subjected to high electron-beam currents, the overall cell-averaged relaxation rate $\langle \Gamma \rangle$ may not be strongly affected even though the local relaxation rate in the target chamber $\Gamma_\mathrm{tc}$ is. As stated earlier, what is important is that the polarized gas in the target chamber is replenished at a rate that is much faster than the rate at which the gas is depolarized in the target chamber. 

Looking at specific examples, we find that the targets used during E142 at SLAC had polarization gradients of just over 1\% with no beam current, and would hypothetically have had polarization gradients of around 13\% at $100\,\mu\rm A$.  For the targets used to measure $G_E^n$ at JLab, the gradients were worse, a little over 6\% with no beam current, and we project they would be about 21\% at $100\,\mu\rm A$.  The larger gradients in the case of the $G_E^n$ cells were mostly due to faster intrinsic cell-relaxation rates, but also to having longer transfer tubes, something that is difficult to avoid with a cell that has a larger overall size. Shortly we will return to the question of what can be done to better minimize polarization gradients  when using a single transfer tube.

\subsubsection{Quantifying polarization gradients}
Regardless of the magnitude of the polarization gradient, it is also difficult to quantify accurately because of uncertainty in the quantity $\Gamma_\mathrm{tc}$.  When characterizing a target cell, the quantity that is most straightforward to measure is the cell-averaged room-temperature spin-relaxation rate $\langle\Gamma\rangle$, which is typically due to three primary contributions:
\begin{equation}
\langle \Gamma \rangle = \Gamma^{w} + \Gamma^{d} + \Gamma^{b}
\label{eq:contributions1}
\end{equation}
where $ \Gamma^w$ is the spin relaxation rate due to wall collisions,  $\Gamma^{d}$ is spin relaxation due to dipolar interactions during $\rm ^3He-{^3He}$ collisions and  $\Gamma^{b}$ is the spin relaxation due to the electron beam, which can be taken to be zero if the cell's spin-relaxation rate is measured in the absence of an electron beam.  Here we ignore relaxation due to magnetic field inhomogeneities which with some care can be made quite small.   The wall relaxation rate will be the sum of the wall relaxation rates in the target and pumping chambers respectively, weighted by the fraction of $^3$He atoms that are in each chamber:
\begin{equation}
\Gamma^w = f_\mathrm{tc} \Gamma_\mathrm{tc}^w + f_\mathrm{pc} \Gamma_\mathrm{pc}^w
\label{eq:contributions2}
\end{equation}
For the purposes of this discussion, it is convenient to introduce a parameter R, representing the ratio of $\Gamma_\mathrm{tc}^w$ to $\Gamma_\mathrm{pc}^w$, so that
\begin{equation}
\Gamma_\mathrm{tc}^w = R\,\Gamma_\mathrm{pc}^w\ \ .
\label{eq:contributions3}
\end{equation}
From Eqns.~\ref{eq:contributions1}--\ref{eq:contributions3}, taking $\Gamma^b=0$, we find an expression for $\Gamma_\mathrm{tc}^w$:
\begin{equation}
\Gamma_\mathrm{tc}^w = {{R(\langle \Gamma \rangle - \Gamma^d)}\over{f_\mathrm{tc}\,R + f_\mathrm{pc}}}
\label{eq:contributions4}
\end{equation}
Unfortunately, we have no direct measurement of $R$, and wall-relaxation rates are notoriously variable.
One plausible assumption is that $\Gamma_\mathrm{tc}^w = \Gamma_\mathrm{pc}^w$, in which case $R=1$, and  Eqn.~\ref{eq:contributions4} simplifies to $\Gamma_\mathrm{tc}^w = \Gamma_\mathrm{pc}^w = \langle \Gamma \rangle - \Gamma^{d}$.  This is, in fact, the assumption that has been made in polarized $^3$He experiments (those using the basic design shown in Fig.~\ref{fig:genstyle}) prior the aforementioned $G_E^n$ experiment. Several authors have shown, however, that for uniform wall relaxation per unit area, overall wall relaxation is proportional to the surface-to-volume (S/V) ratio of the vessel containing the gas~\cite{PhysRev.115.1478, fit69}.  This dependence was recently exploited by Anger {\it et al.} who successfully constructed storage cells for polarized $^{129}$Xe with unusually long, perhaps unprecedented, relaxation times~\cite{ang08}.  If wall relaxation were uniform throughout a target, we would expect $R$ to be the ratio of the S/V ratios of the pumping and target chambers respectively, a quantity we will refer to here as $R_\mathrm{max}$.  A conservative approach, then, might be to take $R=(1+R_\mathrm{max})/2$, with an error on $\Delta$  that includes the full range of $1<R<R_\mathrm{max}$. 

The uncertainties in $\Gamma_\mathrm{tc}$, and consequently $\Delta$, can in some situations translate into a  systematic uncertainty in polarimetry.  One of the best techniques for determining the absolute polarization of $^3$He is the method of measuring shifts in the electron paramagnetic resonance frequencies of the alkali-metal atoms due to the effective magnetic field caused by the polarized $^3$He~\cite{romalis1998}.  This can only be performed in the pumping chamber (where significant alkali-metal vapor is present), despite the fact that the quantity of interest in an electron scattering experiment is the polarization in the target chamber.  Thus, it is often the case that NMR measurements are made directly on the target chamber, but are calibrated against frequency shift measurements in the pumping chamber.  When this is done, the calibration requires a knowledge of $\Delta$.   For the case of the targets used in the $G_E^n$ experiment discussed earlier, the uncertainty in $\Delta$, following the prescription  described earlier, translated into a 1--2\% (relative) uncertainty in polarimetery.   While not catastrophic, such systematic uncertainties would be nice to avoid.

\subsubsection{Limitations from using a single transfer tube}

Two points emerge regarding polarization gradients:  1) they limit target polarization at high beam currents, and 2) they are somewhat uncertain in their size, which in some circumstances, results in systematic uncertainties in polarimetry.   If we want to make polarization gradients smaller while retaining the basic cell design illustrated in Fig.~\ref{fig:genstyle}, there are two things that can be considered: increasing the cross sectional area of the transfer tube and decreasing the length of the transfer tube.  We note that with the large-volume cells that are more resistant to beam current, design constraints make it difficult to significantly shorten the transfer tube, and they are currently at most 50\% longer than the shortest lengths used in early two-chambered targets such as those employed during E142.  We will thus focus here on the cross sectional area.

To better understand the limitations of single transfer-tube configurations for future targets, we consider a hypothetical target design for an approved experiment at JLab (E12-06-016) that will measure $G_E^n$ up to $Q^2=\rm 10\,GeV^2$~\cite{GEN2prop}. The experiment will run at $60\,\mu\rm A$ with a target length of 60~cm instead of 40~cm, a luminosity equivalent to running $90\,\mu\rm A$ into the target of Fig.~\ref{fig:genstyle}.  Here we will assume a single transfer tube is employed instead of the convection-based design that is actually specified in the proposal.

The single best way to make a target less susceptible to beam current is to make the cell larger and use more lasers, thus making relaxation from the beam a smaller perturbation to the target as a whole.  The proposal for E12-06-016 calls for a target containing roughly 7 STP liters of gas, in contrast to the roughly 3 STP liters contained in the  targets illustrated by Fig.~\ref{fig:genstyle}.   Using Fig.~\ref{fig:genstyle} as a starting point, we will assume a transfer tube diameter equal to the diameter of the target chamber (about as large as is practically realizable).  We further assume the transfer tube has the same length as before, and that the pumping chamber is larger (around 12.5~cm) so as to satisfy the criterion of having the target contain 7 STP liters of gas.  Such a cell would have a polarization gradient of around 12\% in $60\,\mu\rm A$ of beam. If the cell-averaged polarization were around 70\% with no beam (more on this in section IV), and 62\% in beam, the polarization in the target chamber would be about 55\%.  In addition to the reduction of the in-beam polarization, the gradients would also have the potential of introducing uncertainties in polarimetry at the level of 2--4\%.  In short, this hypothetical design falls well short of optimized performance.  
And while we have assumed here the same transfer-tube length as is shown in Fig. 1, as will be discussed more in section IV, there are compelling reasons to increase the distance between the pumping and target chambers.  This would naturally require a longer transfer tube, something that would further aggravate the problem of polarization gradients.

\section{Convection Driven Cells}
\label{section:CDC}

We describe next a variant of the target cell geometry depicted in Figure 1. There are still two chambers, a pumping chamber and a target chamber, but the two chambers are connected by two transfer tubes instead of one.   With this design, it is possible to induce convection, thus causing rapid transfer of gas between the two chambers.  Furthermore, all that is required to induce convection is to maintain a temperature differential between the vertical segments of the two transfer tubes.  By controlling the temperature differential, the speed of the convection can be adjusted.  With rapid mixing of gas between the two chambers, the aforementioned polarization gradients can be made negligible, even if the distance between the pumping and target chambers is substantially increased.

\subsection{Experimental Setup}

\begin{figure}[hb]
\begin{center}
\includegraphics[width = 8.6 cm]{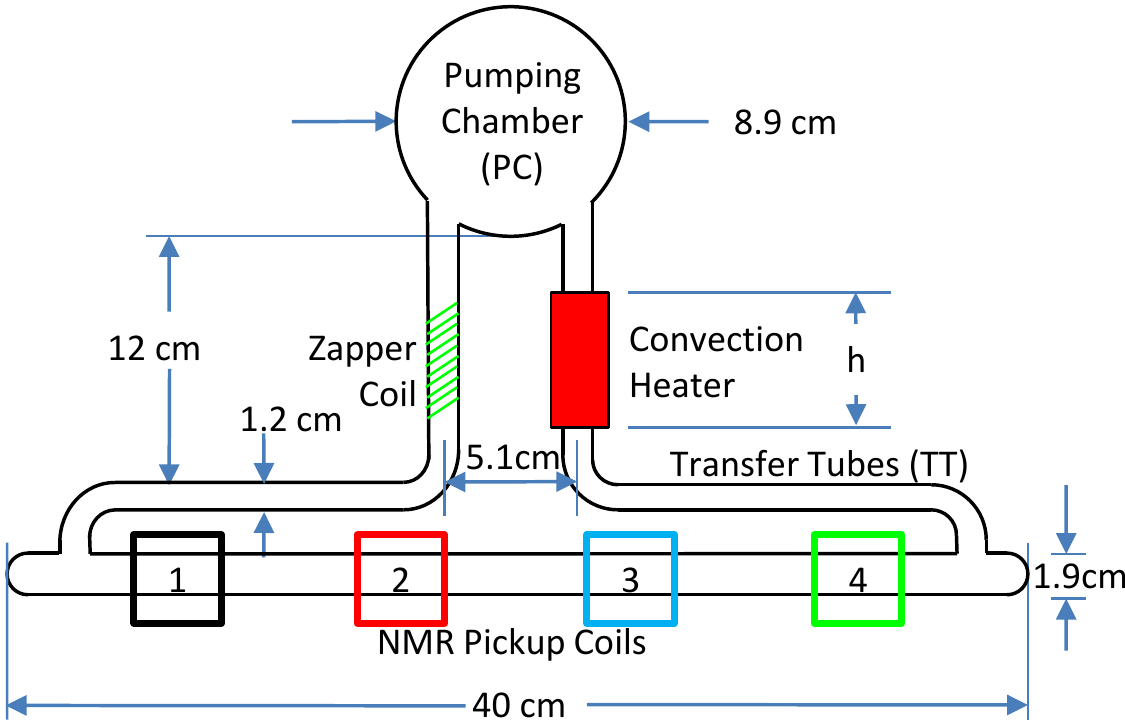}
\caption{Prototype convection-based target cell.  The pumping chamber is placed inside of an optical pumping oven.  The right transfer tube is heated while the left transfer tube is kept at room temperature.  The two transfer tubes have different densities which creates a counter-clockwise convection current in the cell.  The zapper coil is used to depolarize a slug of gas.  This slug is then monitored as it travels through the pickup coils on the target chamber.  We note that a small horizontal portion of the transfer tube was also heated, but for clarity is not shown, since it did not contribute to driving convection.}
\label{fig:convectionstyle}
\end{center}
\end{figure}

To demonstrate the feasibility of convection-driven polarized $^3$He target cells, we have constructed a prototype with the geometry and dimensions illustrated in Fig.~\ref{fig:convectionstyle}.   The cell was constructed entirely out of aluminosilicate glass (GE 180), and was sealed after being filled with 7.2 amg of $^3$He and a 0.11 amg of N$_2$.  The pumping chamber also contained several tens of milligrams of a hybrid mixture of potassium and rubidium. 

The $^3$He gas in the new prototype was polarized in the same manner as in our other target cells.  The pumping chamber was surrounded by a forced-hot-air oven,  constructed largely out of ceramic and glass.  While polarizing the $^3$He, the oven was maintained at temperatures that were typically between $\rm 200-235^\circ$C, resulting in a vapor pressure of alkali-metal atoms corresponding to a number density on the order of $\rm 10^{15}cm^{-3}$.  The rubidium atoms were optically pumped using laser light from high-power diode-laser arrays with a wavelength of 795~nm, and 
as described in Refs.~\cite{happer2001} and \cite{babcock2003},
quickly shared their polarization with the potassium atoms that were also present.  Subsequent spin-exchange collisions with the $^3$He atoms  resulted in the buildup of substantial nuclear polarization.

The temperature differential used to induce convection was maintained by using a second forced-hot-air ``convection heater",  installed on one of the transfer tubes as illustrated on Fig.~\ref{fig:convectionstyle} (we note that, in reality, the convection heater also covered much of the horizontal portion of the transfer tube, but that this portion contributed negligibly to the speed of the convection).  With a portion of one of the transfer tubes at an elevated temperature, the gas contained therein had a lower density than the corresponding gas in the other transfer tube, and thus experienced a small buoyant force which induced convection.  By controlling the temperature of the convection heater, the gas flow could be controlled in a stable and reproducible fashion.

\begin{figure}[htbp]
 \begin{center}
\includegraphics[width = 8.6 cm]{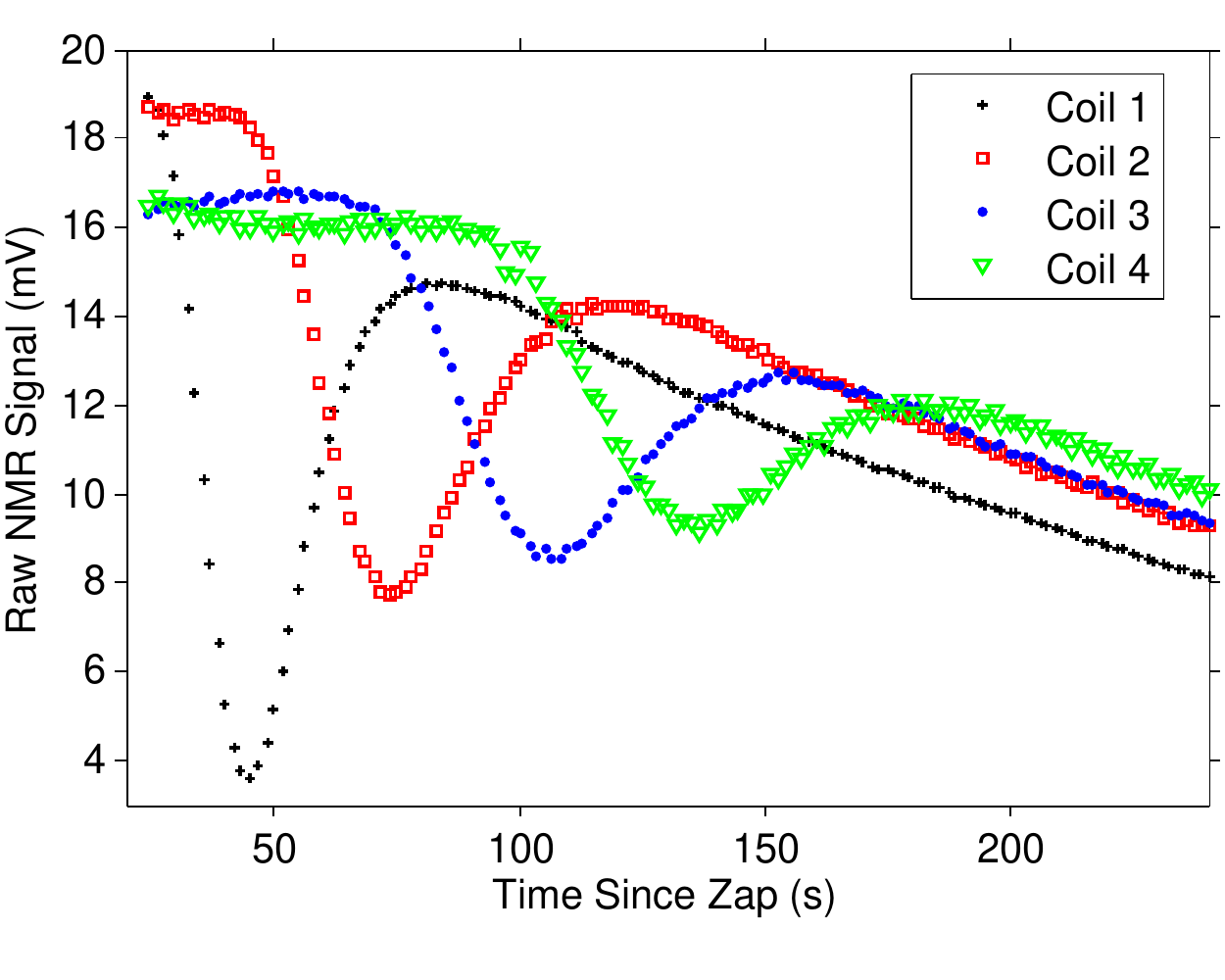}
 \end{center}
 \vskip -.3truein
 \caption{Data used to visualize gas flow are shown in which the NMR signal from four pickup coils are plotted versus time.  The oven temperature was 215$^\circ$C, and the transfer tubes were 24$^\circ$C and 50$^\circ$C respectively.  The data indicate a gas flow velocity of $20\,$cm/min in the target chamber.}
\label{fig:zapper}
 \end{figure}

The flow of the gas was monitored  using an NMR tagging technique.   A  ``slug" of gas  within a small section of one of the transfer tubes was depolarized by subjecting it to a pulse of RF tuned to the Larmor frequency of the $^3$He nuclei.  The RF was delivered  using a small coil (labeled in Fig.~\ref{fig:convectionstyle} as the ``Zapper coil") wrapped directly around one of the transfer tubes.   NMR signals were then detected at each of four locations along the target chamber as indicated on Fig.~\ref{fig:convectionstyle} using small ``pickup coils".  The movement of the slug of gas could then be tracked by monitoring NMR signals from each of the four pickup coils.   Signals from the pickup coils were obtained once every two seconds using the NMR technique of adiabatic fast passage (AFP)\cite{Abragam}.  Representative examples of such signals are plotted in Fig.~\ref{fig:zapper} as a function of time.   Time zero in this plot corresponds to the moment when a slug of gas was tagged.

It is readily apparent from Fig.~\ref{fig:zapper} that a transient dip occurs in each of the signals from the four pickup coils.  This dip corresponds to the passage of the depolarized slug of gas, and it can be seen that the transient occurs at successively later times for each of coils 1--4.   Given the known positions of the regularly spaced pickup coils, the difference in time between the transients associated with each of the four pickup coils provides a measure of the speed with which the tagged slug of gas was moving.  The measurement illustrated in Fig.~\ref{fig:zapper} corresponds to a target-chamber gas velocity of $\rm 20\,cm/minute$.

Also apparent from Fig.~\ref{fig:zapper} is the fact that each successive dip becomes wider and more shallow. This is due in part to the fact that the gas flow within the cell is characterized by the classic parabolic Hagen-Pouiselle velocity distribution.  That is, the velocity as a function of the distance $r$ from the middle of the tube has the functional form  $v(r) = v_\mathrm{max}(1-r^2/R^2)$.  Hence, the slug of gas, which is initially fairly localized to the region around the ``zapper coil", becomes increasingly spread out as it moves through the target chamber.  Additional spreading also occurs because of diffusion.  We note also that the NMR signal decreases as a function of time.  This is due largely to polarization losses that occur with each AFP measurement.  While the loss from each individual measurement is quite small (on the order of 1\%), the accumulation of many such losses is quite substantial.  We note that in the normal operation of a polarized $^3$He target measurements are typically made not every two seconds, but rather every two to four hours.

\subsection{Temperature Dependence of the Gas Velocity}

Hagen-Pouiselle flow occurs whenever a pressure differential between two ends of a pipe causes the laminar flow of a viscous gas or fluid~\cite{dodgeNthompson,tritton}.  In equilibrium, the driving force from the pressure differential $F_\mathrm{driving}$ must be equal to the retarding force $F_\mathrm{retarding}$ from the viscosity:
\begin{equation}
F_\mathrm{driving} = F_\mathrm{retarding} \ \  .
\label{eqn:F}
\end{equation}
For the case of a pipe that is circular in cross section, Eqn.~\ref{eqn:F} must be satisfied for each annular ring of fluid of thickness $dr$, a condition which leads to the equation  
\begin{equation}
\Delta P 2\pi rdr = -2\pi\eta l \frac{d}{dr}\left(r\frac{dv}{dr}\right)dr \ \ ,
\label{eqn:F2}
\end{equation}
where $\Delta P$ is the pressure differential, $\eta$ is the viscosity of the fluid and $l$ is the length of the pipe.  Imposing the boundary condition that the velocity of flow must go to zero at the perimeter of the pipe, the solution to this differential equation is
\begin{equation}
v(r) = \frac{1}{4}\frac{\Delta P(r^2-R^2)}{\eta l}
\label{eqn:vr}
\end{equation}
where  $R$ is the radius of the pipe.  It is the velocity distribution given by Eqn.~\ref{eqn:vr} that is often referred to as Hagen-Pouiselle Flow.

In the case of our convection cell the driving force is due to the small buoyancy force that results from maintaining a vertical portion of one transfer tube at a higher temperature than the corresponding section of the other transfer tube:
\begin{equation}
F_\mathrm{buoyancy} = \Delta\rho\,V_\mathrm{t}\,g
\end{equation}
where $V_\mathrm{t}$ is the volume of the vertical portion of the transfer tube that is being heated, $\Delta\rho$ is the difference between the average densities of the heated and unheated portions of the transfer tubes, and $g$ is the acceleration due to gravity.  We can express  $\Delta\rho$ as follows:
\begin{equation}
\Delta\rho = \rho\,T_\mathrm{C}\,\left({1\over{T_\mathrm{C}}} - {1\over{T_\mathrm{H}}}\right)
\label{eq:del_rho}
\end{equation}
where $ \rho$ is the density of the gas in those portions of the cell that are not heated, $T_\mathrm{C}$ is the temperature of those portions of the cell that are not heated, and $T_\mathrm{H}$ is the temperature of the portion of the transfer tube that is being heated (both temperatures are in Kelvin).
For the case being discussed here, the pressure differential that appears in the left-hand side of Eqn.~\ref{eqn:F2} is thus given by 
\begin{equation}
\Delta P = F_\mathrm{buoyancy}/A_\mathrm{t} =  \Delta\rho\,h\,g, 
\label{eq:del_p}
\end{equation}
where $A_\mathrm{t}$ is the cross sectional area of the transfer tube, and $h$ is the length of the portion of the transfer tube that is heated.  

If our convection cell could be treated as a long straight tube, we could simply substitute Eqns.~\ref{eq:del_rho} and \ref{eq:del_p} into Eqn.~\ref{eqn:vr} to obtain an expression for the velocity.  Our convection cells are more complex, however which complicates the expression that appears on the right-hand side of Eqn.~\ref{eqn:F2}.  There are multiple sections of tubing, each with its own radius, as well as bends etc.   The velocity of the gas will be different in each section, and even the viscosity will be different depending on the temperature.  Luckily, however, as will be shown in the appendix, the continuity equation ensures that the velocity in each section is related in a simple linear fashion to the velocity in the other sections, so it is still possible to solve Eqn.~\ref{eqn:F2} exactly.  In essence, the quantity $\eta l$ that appears on the right-hand side of Eqn.~\ref{eqn:F2} must be replaced by a single quantity $k_{\eta\mathcal{L}}$, which still has the dimensions of viscosity times length, but incorporates the full complexity of the cell.   The solution can accordingly be written in the form
\begin{equation}
v(r,T_\mathrm{H}) = \frac{(r^2-R^2)}{4\,k_{\eta\mathcal{L}}}\rho\,h\,g\,T_\mathrm{C}\,\left({1\over{T_\mathrm{C}}} - {1\over{T_\mathrm{H}}}\right)\ \ ,
\label{eqn:vel_convec}
\end{equation}
where $r$ is the radial coordinate in the target chamber, and $R$ is the radius of the target chamber.
The temperature dependence of $v$ is largely dominated by the factor $({1\over{T_\mathrm{C}}} - {1\over{T_\mathrm{H}}})$.  The quantity
$k_{\eta\mathcal{L}}$, however, is also dependent on temperature, although for the range of values of $T_\mathrm{H}$ that we consider, the temperature dependence of $k_{\eta\mathcal{L}}$ on $T_\mathrm{H}$ is relatively weak. 

It is important to understand the relationship between the observed velocity of the gas, $v^\mathrm{obs}$, as indicated by data of the sort shown in Fig.~\ref{fig:zapper}, and the velocity distribution given by  Eqn.~\ref{eqn:vel_convec}. The natural way to compute $v^\mathrm{obs}$ is by taking the physical separation of adjacent pickup coils, and dividing by the separation in time between the minima of the corresponding transients.  If we consider the limit in which the distance between the zapper coil and the pickup coils is long compared to the length of the zapper coil itself, it is straightforward to show that, to a good approximation, the above method of computing $v^\mathrm{obs}$ corresponds to the maximum value of the velocity given by Eqn.~\ref{eqn:vel_convec}, $v^\mathrm{max}$, which results from setting $r=0$.  
To a good approximation, we can express $v^\mathrm{max}$ in the simplified form
\begin{equation}
v^\mathrm{max} = \frac{\mathcal{A}}{1 + \beta_1 \Delta T}\left(\frac{1}{T_\mathrm{C}} - \frac{1}{T_\mathrm{H}}\right)\ ,
\label{eq:temp_fit}
\end{equation}
where all quantities not dependent on $T_\mathrm{H}$ have been absorbed into the constant $\mathcal{A}$, and the temperature dependence of $k_{\eta\mathcal{L}}$ is accounted for by the factor ${1 + \beta_1 \Delta T}$, where $\Delta T=T_\mathrm{H}-T_\mathrm{C}$.
For the conditions we have considered, $\beta_1$ is on the order of $10^{-3}/^\circ \rm$ C.

\begin{figure}[htbp]
 \begin{center}
\includegraphics[width = 8.6 cm]{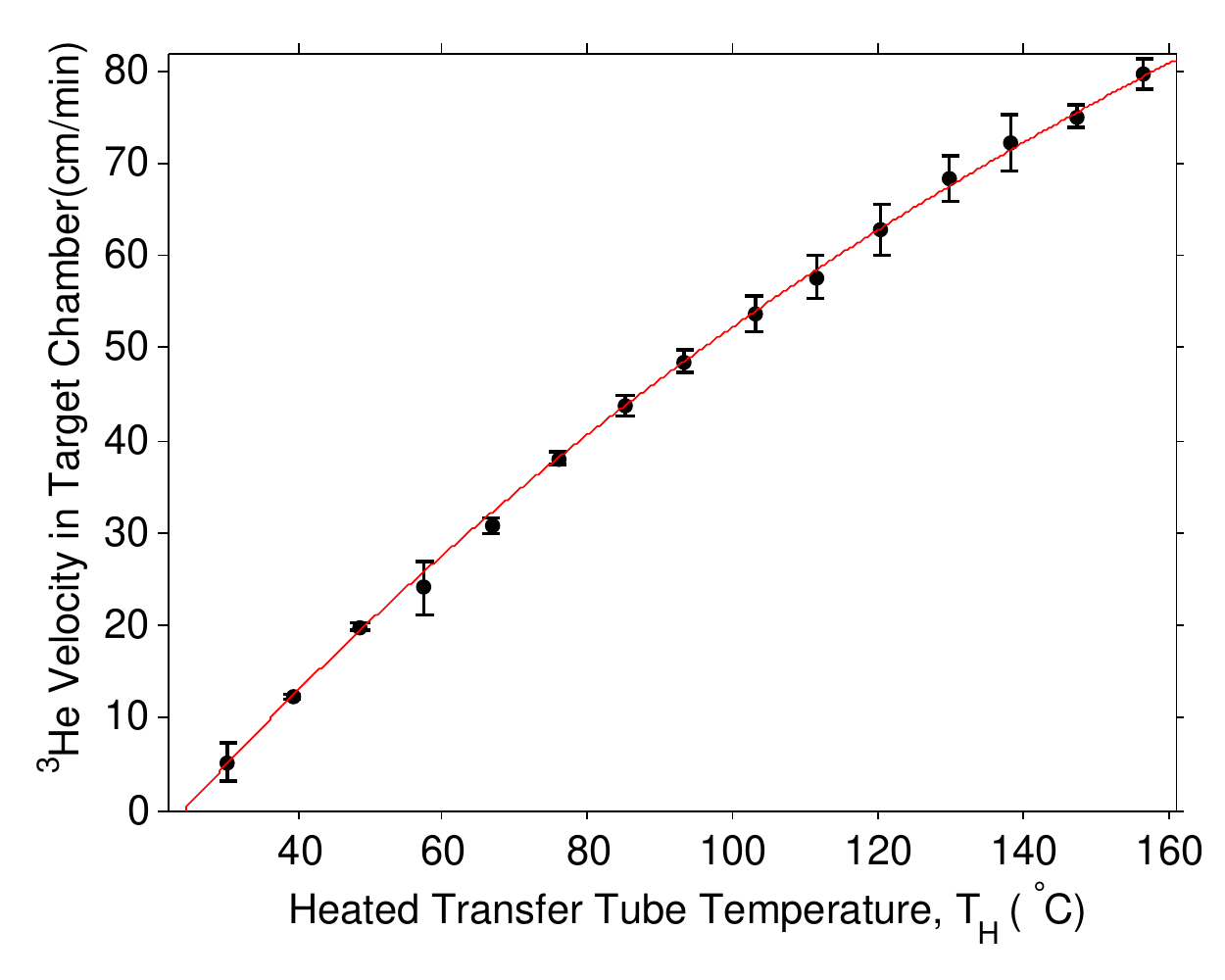}
 \end{center}
 \vskip -0.3truein
 \caption{The velocity of the gas in the target chamber of the convection-based cell is shown as a function of the temperature T of the heated transfer tube.  The oven temperature was 215$^\circ$C and the unheated transfer tube was at 24$^\circ$C.}
\label{fig:convectiondata}
 \end{figure}

In Fig.~\ref{fig:convectiondata} we have plotted $v^\mathrm{obs}$ as a function of $T_\mathrm{H}$  for a range of temperatures.  For each point,  the velocity was computed using data such as those shown in Fig.~\ref{fig:zapper}.
We also show in Fig.~\ref{fig:convectiondata}  with a solid black line a fit of the data to a function of the form of 
Eqn.~\ref{eq:temp_fit}.
The quality of the fit is clearly quite good, and yields the 
values  $\mathcal{A} = 7.47(22) \times 10^4$K$\cdot$cm/min, $T_\mathrm{cold} = 24.3(8)$, and $\beta_1 = -0.2(2)\times 10^{-3}/^\circ \rm$ C. 
For comparison, using our best knowledge of the cell geometry and densities, with $T_\mathrm{cold} = 24.5^\circ$C, we compute $\mathcal{A} = 9.14 \times 10^4$K$\cdot$cm/min and $\beta_1 = 1.03\times 10^{-3}/^\circ \rm$ C (see Appendix A).  Given the uncertainties of some of the quantities with which we are working, particularly in describing the retarding forces associated with our relatively complicated cell geometry, this agreement is quite reasonable.  More importantly, the agreement is more than sufficient to suggest that we have an acceptable quantitative understanding of the parameters influencing the convective flow from a practical perspective.

\subsection{Convection Transfer Rates and the Elimination of Polarization Gradients}

Ultimately, the value of a convection-driven target cell is measured by the degree to which polarization gradients can be avoided between the pumping chamber and the target chamber.   It is critical that, as gas is depolarized by an electron beam, freshly polarized gas is delivered from the pumping chamber sufficiently quickly.  Ideally, one would like the ratio of the polarizations of the two chambers to be as close to unity as possible.  Fortunately, when convection is used to transfer gas between chambers, the polarization gradient is suppressed because the transfer rates are quite high.  

For the range of velocities measured in Fig.~\ref{fig:convectiondata}, we find values for $d_\mathrm{tc}$ in the range of 4.9 -- 81 hrs$^{-1}$.  Even with relatively fast relaxation in the target chamber (for example, consider $\Gamma_\mathrm{tc} = 1/10\,$hrs$^{-1}$), the polarization gradient (Eqn.~\ref{eq:delta}) will be very small in the presence of such fast transfer rates ($\Delta \leq 0.02$, or $P_\mathrm{tc}/P_\mathrm{pc} \geq 0.98$).

\begin{figure}[htbp]
 \begin{center}
 \includegraphics[width = 8.6 cm]{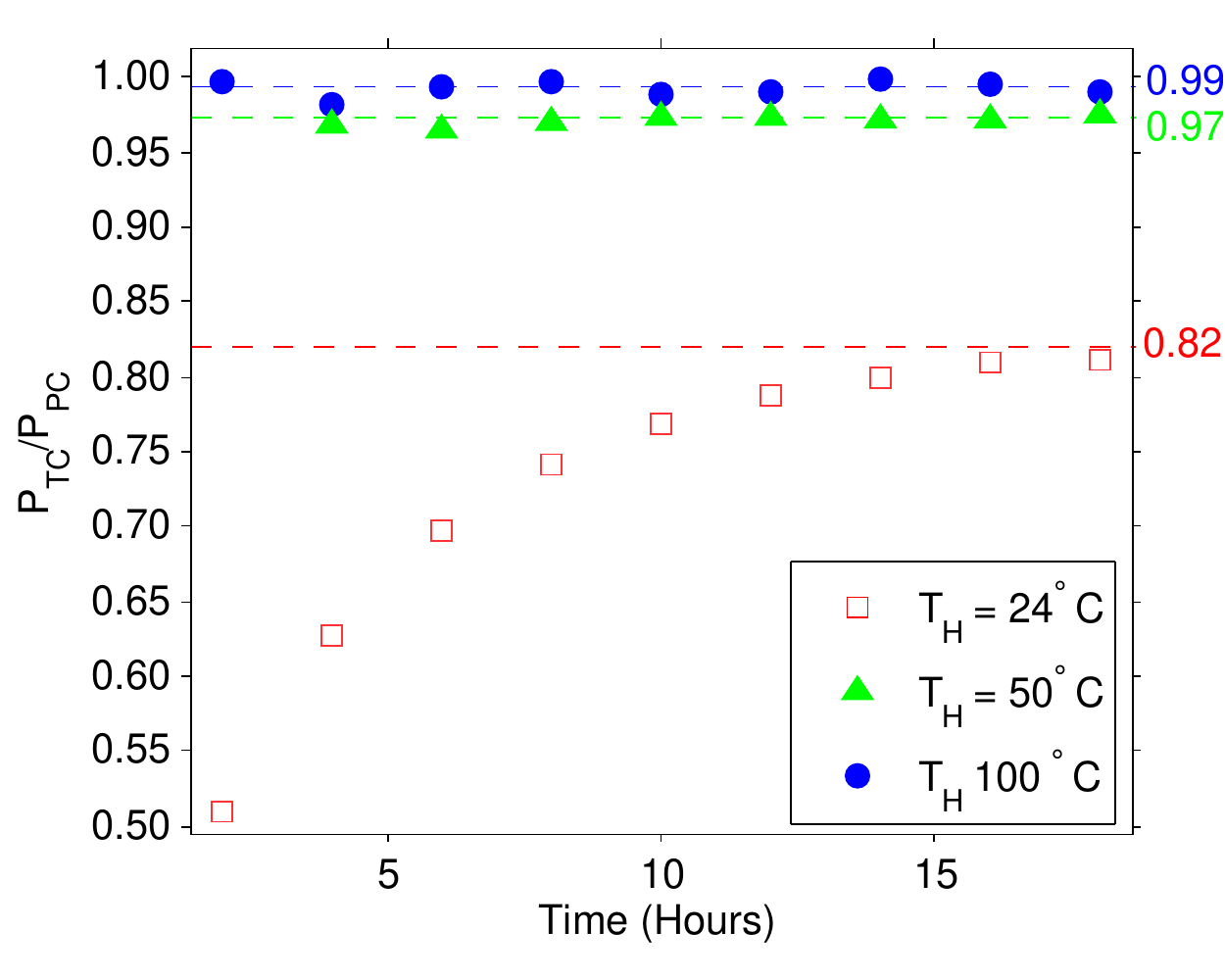}
 \end{center}
 \vskip -0.3truein
 \caption{The ratio of the polarizations of the target chamber and the pumping chamber, P$_\mathrm{tc}/$P$_\mathrm{pc}$, is shown versus time for 3 spin-ups corresponding to different convection velocities.}
\label{fig:ConvectionRatio}
 \end{figure}

In Fig.~\ref{fig:brady}, the $^3$He polarization in both chambers of a traditional target cell is plotted as a function of time.  The ratio of target chamber to pumping chamber polarization for this plot begins at zero, and gradually climbs to a value significantly less than unity, reflecting a substantial polarization gradient.  In Fig.~\ref{fig:ConvectionRatio}, we plot this ratio for a convection-driven cell for three different operating conditions.  In all cases, the temperature of the oven was held at $\rm 215\,^{\circ}C$.  The three curves correspond to different temperatures $T_\mathrm{H}$ of the heated transfer tube.  For the data shown with the open squares, the transfer-tube set temperature was $\rm 24\,^{\circ}C$, the same as the other (unheated) transfer tube.  This case corresponds to no driven convection, and is associated with a polarization gradient of 8\%.   For the data shown with the filled triangles and the filled circles, the set temperatures were $\rm 50\,^{\circ}C$ and $\rm 100^{\circ}C$ respectively.  These two conditions corresponded to target-chamber gas velocities of approximately 19.9 cm/min and 48.5 cm/min.  In both of these cases, the ratio of the polarizations of the target chamber and the pumping chamber quickly reached a value approaching unity.  It is notable that there is very little difference between these last two curves despite substantially different gas velocities.  In short, as soon as convection rather than diffusion is responsible for the gas transfer between the two chambers, polarization is relatively uniform throughout the cell.

\section{Outlook For Future Targets}

With advances in SEOP, it is now possible to run polarized $^3$He targets at significantly higher luminosities.   Using a single transfer tube, however, even currently approved experiments would suffer polarization gradients of 12\% or worse, and subsequent experiments at even higher luminosities would be even more limited.  The work described here demonstrates that with only minor changes to current designs, convection-based cells can be built in which polarization gradients are extremely small.  

In many ways, however, the most compelling reason to adopt convection-based cells is the enormous flexibility that is gained in being able to choose the distance between the pumping and target chambers.   Experience has shown that the probability of a cell rupturing goes up markedly after 4--6 weeks exposure to something on the order of  $12\,\mu\rm A$ of beam.  Most commonly, the part of the cell that breaks is the pumping chamber.  Stress in a sphere goes up linearly with the radius. This makes the relatively large pumping chamber particularly vulnerable to radiation damage.   With convection-based cells the distance between the pumping chamber and the target chamber can be greatly increased with minimal implications for performance.  Among other things, it becomes practical to incorporate radiation shielding for the pumping chamber.

There are at least two additional  reasons that flexibility in the configuration of target cells is highly desirable.  As discussed earlier, the primary way to make targets more tolerant of beam is to make them larger.  The more gas that is being polarized per unit time, the less relevant it is that some small portion of that gas is being depolarized by the beam.  It is not practical, however, to make the pumping chamber arbitrarily large.  As the stress in the glass walls increases, the cell becomes more prone to rupturing, even if the wall thickness is increased.  Furthermore, the intensity of the optical pumping radiation goes like total power over the square of the radius while the volume goes like the cube of the radius.  Hence, since the required laser power scales with volume, the required intensity of light scales linearly with radius.   There are already indications that we are near a threshold at which laser intensity damages the cells from the perspective of spin relaxation.   If multiple pumping chambers were used instead of one, it would become possible to limit both laser intensity and the size of each individual pumping chamber.  If diffusion were the only mechanism for moving polarized gas inside the target, the implementation of multiple pumping chambers would be quite challenging.

\begin{figure}[htbp]
 \begin{center}
 \includegraphics[width = 8.3 cm]{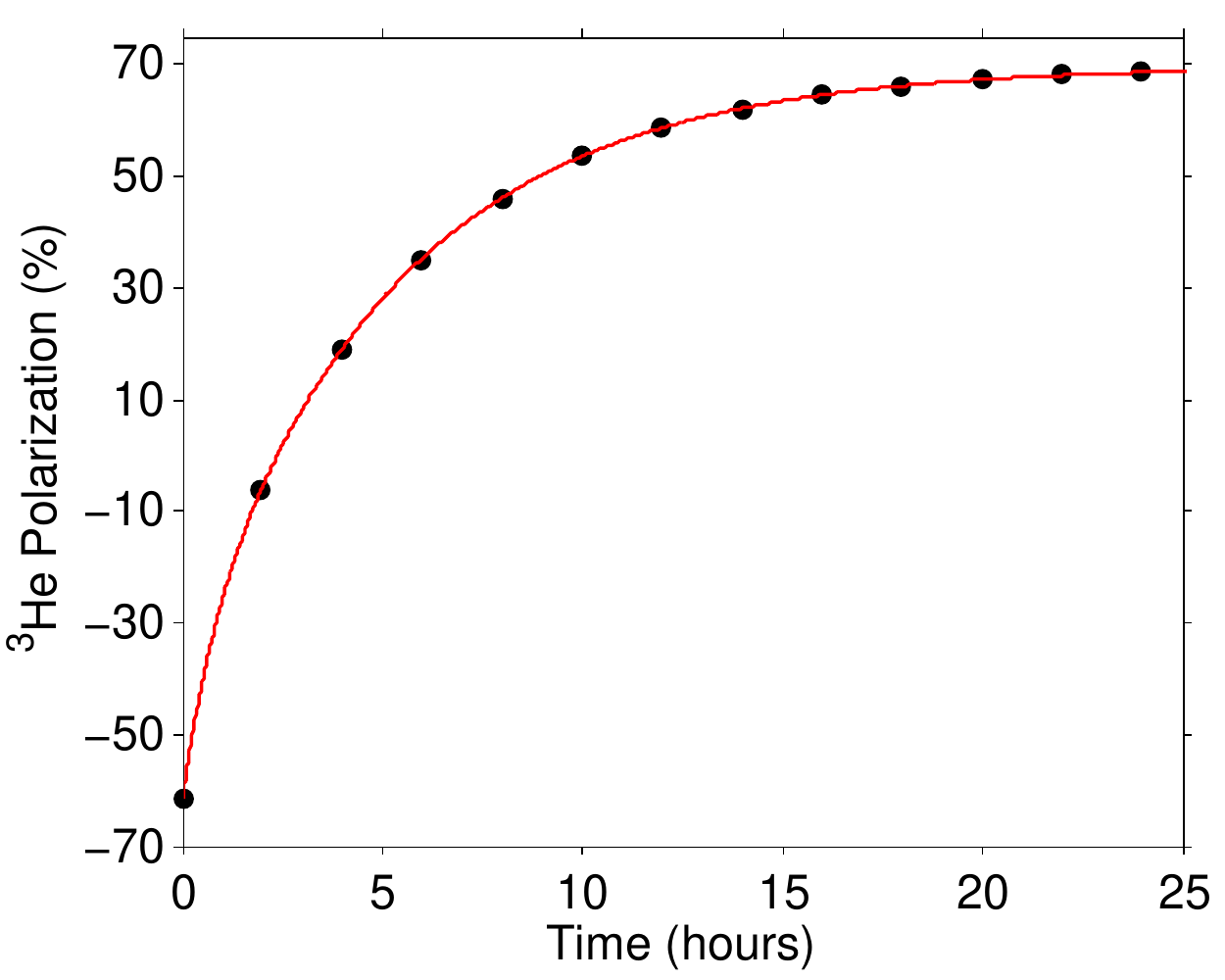}
 \end{center}
 \vskip -0.2truein
 \caption{Shown is polarization as a function of time for the cell ``Brady".  A polarization of 70\% was ultimately reached.  The set of tests giving rise to these data established the highest polarizations observed to date among large  polarized $^3$He target cells intended for electron scattering.}
\label{fig:bradySpinup}
 \end{figure}

We close by citing a concrete example that underscores the desirability of adopting convection-based target cells.  We show in Fig.~\ref{fig:bradySpinup} 
results from a bench test of the cell ``Brady" (discussed earlier) in which the primary goal was to achieve the highest polarization possible.   Brady, like the the targets used for the $G_E^n$ measurements~\cite{PhysRevLett.105.262302}, contained an alkali-hybrid mixture of K and Rb.  Unlike during the $G_E^n$ experiment, however, the results shown in  Fig.~\ref{fig:bradySpinup} were obtained using diode-laser arrays that were spectrally-narrowed such that their line widths were about ten times narrower than their broadband counterparts.   The polarization in this test saturated just over 70\%.  Brady was one of two target cells used to collect data on single-spin asymmetries in semi-inclusive deep inelastic scattering, and was similar in design, although slightly smaller, than the configuration shown in Fig.~\ref{fig:genstyle}.  During the experiment, the in-beam polarization of the $^3$He averaged around 55\%~\cite{qia11}, despite the fact that the cell-averaged polarization, averaged over the entire experiment, was over 60\%.  If the targets used for this experiment had had convection-based gas mixing (which was just being explored at the time), significant gains in performance would have resulted.  

In conclusion, with the advances in target performance brought about by a combination of alkali-hybrid spin-exchange optical pumping and spectrally-narrowed diode-laser arrays, the point has been reached when it is desirable to explore target-cell geometries that go beyond that shown in Fig.~1.  We suggest that the convection-based gas mixing demonstrated in this work can play an important role in implementing high-luminosity polarized $^3$He targets in the future.  Not only will the use of convection reduce polarization gradients, it will also make possible greater flexibility in the placement of the pumping chamber with respect to the target chamber.  Among other things, this flexibility will enable the use of radiation shielding, and even the use of multiple pumping chambers.   Convection-based polarized $^3$He target cells open new possibilities that collectively address several of the important challenges facing the next generation of high-luminosity polarized $^3$He targets.

\begin{acknowledgments}
This work was supported by the U.S. Department of Energy under
contract numbers DE-FG02-01ER41168 and DE-AC05-060R23177.
Jefferson Science Associates, LLC, operates Jefferson Lab for the U.S. DOE under U.S. DOE Contract No. DE-AC05-060R23177.  We would also like to thank Michael Souza of Princeton University
for his patience and expert glass blowing.
\end{acknowledgments}

\appendix

\section{Estimate of the magnitude of flow in the convection cells.}

Here we formulate estimates for the parameters that appear in Eqn.~\ref{eq:temp_fit}.

\subsection{The Viscosity of $^3$He}

Since our temperatures are in the classical regime ($T \gg$ 3 K), the viscosity $\eta_{\rm He3}$ of $^3$He  can be calculated from the viscosity $\eta_{\rm He4}$ of $^4$He using~\cite{PhysRev.105.41}:
\begin{eqnarray}
\eta_{\rm He3} &=& \sqrt{\frac{m_{\rm He3}}{m_{\rm He4}}}\eta_{\rm He4} \\
&=& 0.8681~\eta_{\rm He4}\ \ .
\end{eqnarray}
We parameterize the $^4$He viscosity in the range of $0-300^\circ$ C following Kestin et al.~\cite{Kestin},
\begin{equation}
\eta_{\rm He4} = A + B\times T + C\times T^2
\end{equation}
where $T$ has units of $^\circ$ C and
\begin{eqnarray} 
A &=& 18.82(2)~\mu {\rm Pa\cdot s}\ \ , \\
B &=& 0.0456(2)~\mu {\rm Pa\cdot s/^\circ C} \\
{\rm and}\ \ C &=& -13.8(6)\mathrm{p} {\rm Pa\cdot s/(^\circ C)^2}\ \ .
\end{eqnarray}
At $20^\circ C$, $\eta_{\rm He3} = 17.12 \mu$Pa$\cdot$s.

The flow in a pipe is Laminar if the Reynold's number is below 2300~\cite{schiller}.  The Reynolds number is defined as
\begin{equation}
Re = \frac{2R\rho v }{\eta},
\end{equation}
where $\rho$ is the density of the fluid.  A pipe $1.2$cm wide that is filled with 8 amagats $^3$He  will have laminar flow at $20^\circ$ C ($\rm \rho \approx 1 kg\,/m^3$) provided $v \ll 20000$~cm/min.

\subsection{Flow in the Convection Cell}

\begin{figure}[htbp]
 \begin{center}
\includegraphics[width = 8.6 cm]{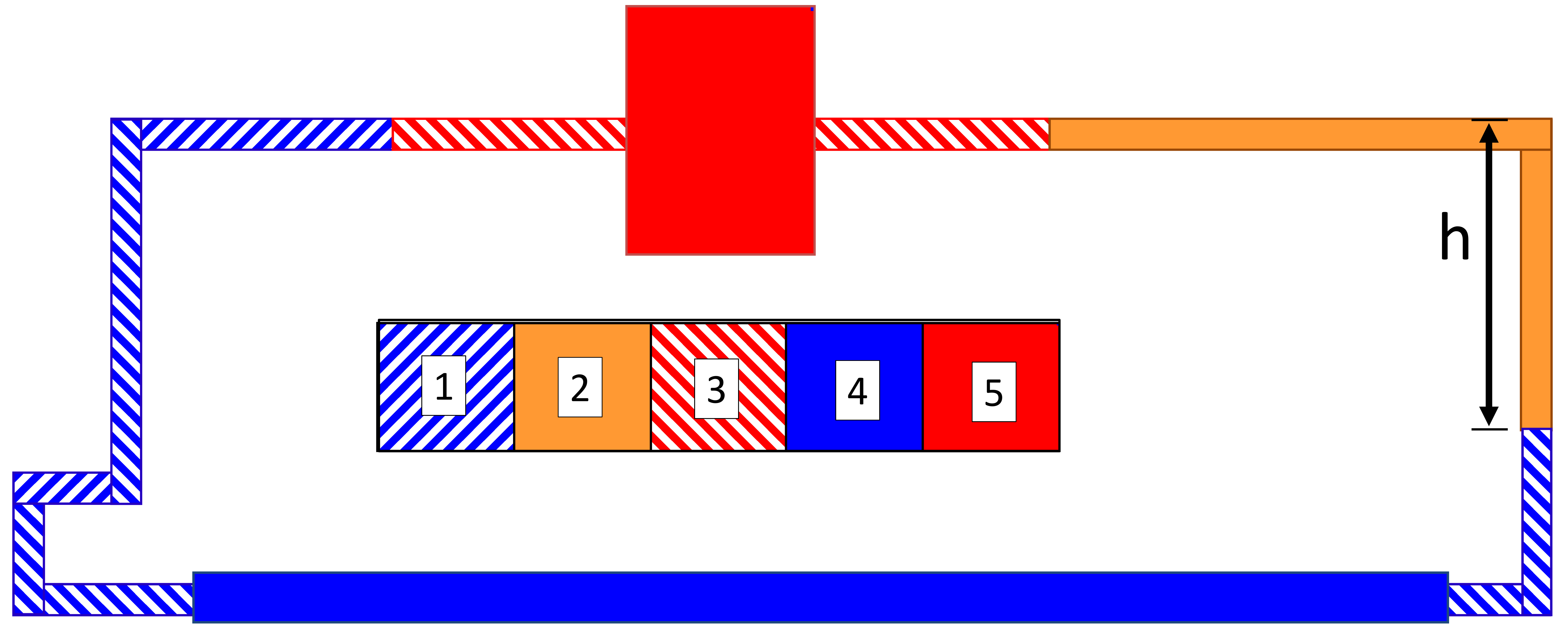}
 \end{center}
 \caption{Diagram indicating the different temperature regions used to describe the convection cell.}
\label{fig:cylinders}
 \end{figure}

The flow in the convection cell arises from a forced density difference between the two transfer tubes -- one tube is maintained at room temperature while the other is heated (see Fig.~\ref{fig:convectionstyle}).  We modeled the convection cell as five contiguous pipes as is illustrated in Fig.~\ref{fig:cylinders}.   Eqn.~\ref{eqn:F2} becomes
\begin{equation}
\Delta \rho gh 2\pi rdr =  -2\pi\displaystyle\sum_{i}^5\eta_\mathrm{i} l_\mathrm{i} \frac{d}{dr_\mathrm{i}}\left(r_\mathrm{i}\frac{dv_\mathrm{i}}{dr_\mathrm{i}}\right)dr_\mathrm{i}
\end{equation}
where $h$ is the vertical length of transfer tube that is held at an elevated temperature.
In this model we approximate the pumping chamber as a cylinder with transfer tubes entering axially, and identify five distinct regions in the cell as is indicated in  Fig.~\ref{fig:cylinders}.  Each region is identified as a pipe of length $l_\mathrm{i}$, radius $R_\mathrm{i}$, cross sectional area $A_\mathrm{i}$ and  temperature $T_\mathrm{i}$.  We further assume that $T_1 = T_4$, $T_3 = T_5$ and $R_1 = R_2 = R_3$.  Finally, we identify $T_\mathrm{C}\equiv T_1$ and $T_\mathrm{H} \equiv T_2$.  We note that both the density and viscosity of the gas are temperature dependent.  

The continuity equation, $\rho_jA_jv_j = \rho_\mathrm{i}A_\mathrm{i}v_\mathrm{i}$ and some distance rescaling provide further simplification:

\begin{eqnarray}
v_\mathrm{i} &=& \frac{\rho_1 R_1^2}{\rho_\mathrm{i} R_\mathrm{i}^2}v_1\ \ , \\
r_\mathrm{i} &=& \frac{R_\mathrm{i}}{R_1}r_1 \\
{\rm and}\ \ \frac{d}{dr_\mathrm{i}} &=& \frac{R_1}{R_\mathrm{i}}\frac{d}{dr_\mathrm{1}}\ \ .
\end{eqnarray}
Since $v_\mathrm{i}$ and $r_\mathrm{i}$ have been expressed in terms of $v_1$ and $r_1$, we'll drop their subscripts.  Finally,
\begin{equation}
v(r) = \frac{1}{4}\frac{\Delta\rho g h\left(r^2-R_1^2\right)}{\eta_1\left(l_1+l_4\frac{R_1^2}{R_4^2}\right) + \eta_2 l_2\frac{\rho_1R_1^2}{\rho_2R_2^2} + \eta_3\frac{\rho_1}{\rho_3}\left(l_3\frac{R_1^2}{R_3^2}+l_5\frac{R_1^2}{R_5^2}\right) }
\label{eqn:pouisellelike1}
\end{equation}
where $v(r)$ is the velocity in region 1.

Eqn.~(\ref{eqn:pouisellelike1}) assumes that there are no ``minor losses" in the system, where a minor loss represents a pressure drop due to a sudden change in flow from a pipe fitting or a pipe expansion or contraction; the term ``minor loss" refers to a loss that is small relative to the overall length of pipe under consideration~\cite{crane,dodgeNthompson}.  In our case, due to the relatively short length of the cell, the minor losses are actually significant.  The actual cell, as illustrated in Fig.~\ref{fig:convectionstyle}, consists of four elbow bends, two tees, and four expansions/contractions.  The retarding forces these losses exert on the gas can be approximated by considering instead an equivalent length of straight pipe.  

The expansions/contractions have a relatively small loss, which is equivalent in magnitude to the loss that would be incurred passing through a pipe of length~\cite{crane}
\begin{equation}
L_\mathrm{equivalent} = \frac{2RK}{f},
\end{equation}
where R is the tube radius, $f$ is the friction factor ($f = 64/Re$ for laminar flow) and $K$ is the fluid-independant resistance coefficient.  For sudden expansions/contractions,
\begin{eqnarray}
K_\mathrm{expansion} &=& \left(1-\frac{r_\mathrm{small}^2}{r_\mathrm{large}^2}\right)^2 \\
{\rm and\ \  } K_\mathrm{contraction} &=& \frac{1}{2}\left(1-\frac{r_\mathrm{small}^2}{r_\mathrm{large}^2}\right)^2 \ .
\end{eqnarray}
Gas flowing through the transfer tube/target chamber junction has $K_\mathrm{contraction} \approx 0.18$, $K_\mathrm{expansion} \approx 0.35$.  At $v=60\,$cm/min ($Re \approx 10$), this gives a negligible $L_\mathrm{equivalent} \approx 0.1$cm -- gas flowing between the pumping chamber and the transfer tube will have an even smaller $L_\mathrm{equivalent}$.

The losses in the bends, however, are much greater.  We model the loss coefficient in the bends using the 3-K method of Darby~\cite{Darby},
\begin{equation}
K = \frac{K_1}{Re} + K_\mathrm{i}\left(1+\frac{K_d}{D^{0.3}}\right)
\end{equation}  
where $D$ is the diameter of the pipe (in inches) and $K_1,K_\mathrm{i},K_d$ are geometry-dependent loss coefficients.  We approximate our glass bends as flanged, welded bends with $r_b/D = 2$ (here, $r_b$ is the radius of the bend); such bends have $K_1 = 800, K_\mathrm{i} = 0.056, K_d = 3.9$.  For laminar flow, the 3-K method gives
\begin{equation}
L_\mathrm{equivalent} = \frac{2R}{64}\left[K_1 + ReK_\mathrm{i}\left(1+\frac{K_d}{D^{0.3}}\right)\right]\ \ .
\end{equation}
We treat the transfer tube/target chamber tee junctions as elbows (effectively ignoring the dead-end branch of the tee).  The system therefore has five bends in temperature-region 1 (which have a total equivalent length of approximately $\rm 63\,cm$) and one bend in temperature-region 2 (which has an equivalent length of approximately $\rm 13\,cm$).
Finally, using Eqn.~\ref{eq:del_rho} for $\Delta \rho$ we find
\begin{equation}
v(r) = \frac{{1\over4}   \left(r^2-R_1^2\right)  \rho h g T_\mathrm{C}\left(\frac{1}{T_\mathrm{C}} - \frac{1}{T_\mathrm{H}}\right)}
{\eta_1\left(l_1'+l_4\frac{R_1^2}{R_4^2}\right) + \eta_2 l_2'\frac{\rho_1R_1^2}{\rho_2R_2^2} + \eta_3\frac{T_3}{T_1}\left(l_3\frac{R_1^2}{R_3^3}+l_5\frac{R_1^2}{R_5^2}\right) }
\label{eqn:pouisellelike}
\end{equation}
where 
\begin{eqnarray}
l_1' &=& l_1 + 5\times \frac{2R_1}{64}\left[K_1 + Re_1K_\mathrm{i}\left(1+\frac{K_d}{D_1^{0.3}}\right)\right]\quad  \\
l_2' &=& l_2 + \frac{2R_2}{64}\left[K_1 + Re_2K_\mathrm{i}\left(1+\frac{K_d}{D_2^{0.3}}\right)\right]\ \ .
\end{eqnarray}
Note that the Reynold's number is dependent on the velocity of the gas.  

Table~\ref{tab:bestguesses} lists values for R and $l$.  Measurements of $R$ (which is the inner diameter of the tube) require a knowledge of the thickness of the glass tube.  We measured the thickness of the glass by observing interference patterns using a scannable single-frequency laser.  Using this information, Eqn.~\ref{eqn:pouisellelike} predicts velocities that agree within 20\% of the measured value (see Fig.~\ref{fig:convectiondata}).

\begin{table}[h!]
\vskip 0.15truein
\begin{center}
\begin{tabular}{|c|c|c|c|c|c|c|c|c|c|c|}
\hline
R$_1$ & R$_2$ & R$_3$ & R$_4$ & R$_5$ & h & l$_1$ & l$_2$ & l$_3$ & l$_4$ & l$_5$\\
\hline
.498 & .521 & .502 & .806 & 4.034 & 4.76 & 24.99 & 20.87 & 15.18 & 40.32 & 5.6 \\
\hline
\end{tabular}
\caption{Best Guesses for Convection Cell Dimensions (cm)}
\end{center}
\label{tab:bestguesses}
\end{table}

The maximum velocity in Eqn.~\ref{eqn:pouisellelike} (corresponding to $r = 0$) can be written as

\begin{equation}
v_\mathrm{max} = \frac{\mathcal{A}^{\prime} \left(\frac{1}{T_{\mathrm{cold}_{\mathrm{tt}}}} - \frac{1}{T_\mathrm{hot_\mathrm{tt}}}\right)}{1 + \beta(\Delta T)}
\end{equation}
where
\begin{eqnarray}
\Delta T &=& T_\mathrm{H}-T_\mathrm{C} \ \ , \\
\mathcal{A}^{\prime} &=& \frac{  R_1^2  \rho h g  T_\mathrm{C} }{4\,k_{\eta \mathcal{L}}} \ \ , \\
\beta &=& \beta_1\Delta T + \beta_2 (\Delta T)^2 + \beta_3 (\Delta T)^3 \ \ , \\
k_{\eta \mathcal{L}} &=&\eta_1\left(l_1'+l_4{{R_1^2}\over{R_4^2}}\right)+\eta_2l_2'\frac{\rho_1R_1^2}{\rho_2R_2^2}  \cr
& & \quad \quad \quad \quad \quad \ \ \, + \eta_3\frac{T_3}{T_1}\left(l_3\frac{R_1^2}{R_3^2}+l_5\frac{R_1^2}{R_5^2}\right), \\
\Delta T &=& T_2 - T_1 \ \ , \\
\beta_1 &=& l_2'{{R_1^2}\over{R_2^2}}[{{\eta_1}\over{T_1}} + 0.8681(B \cr
& & \quad \quad \quad \quad + 2C(T_1 -273))] \ \ , \\
\beta_2 &=& 0.8681 l_2'\frac{R_1^2}{R_2^2}\left[\frac{B + 2C(T_1-273)}{T_1} + C\right],\\
\mathrm{and} \\
\beta_3 &=& 0.8681 l_2'\frac{R_1^2}{R_2^2}\left[\frac{C}{T_1}\right] \ \ .
\end{eqnarray}
In the above equations, all temperatures are in Kelvin.  To describe the velocity in region 4 (the target chamber) instead of in region 1, we use Eqn.~A9 and replace $\mathcal{A}^{\prime}$ with $\mathcal{A}=(\frac{R_1^2}{R_4^2})\mathcal{A}^{\prime}$.  This is the quantity that appears in Eqn.~\ref{eq:temp_fit} that characterizes the magnitude of the gas velocities shown in Fig.~\ref{fig:convectiondata}.


Evaluating the above equations in terms of our best guesses for cell dimensions and temperatures, with $T_\mathrm{C} = 24.5^\circ$C gives $\mathcal{A} = 9.14 \times 10^4$K$\cdot$cm/min, $\beta_1 = 1.03\times 10^{-3}$,  $\beta_2 = 1.26\times 10^{-6}$, and $\beta_3 = -4.26\times 10^{-10}$.  It is clear from these values that it is quite reasonable to neglect the terms involving $\beta_2$ and $\beta_3$, which results in Eqn.~\ref{eq:temp_fit} from section III-B.


\begin{thebibliography}{40}
\expandafter\ifx\csname natexlab\endcsname\relax\def\natexlab#1{#1}\fi
\expandafter\ifx\csname bibnamefont\endcsname\relax
  \def\bibnamefont#1{#1}\fi
\expandafter\ifx\csname bibfnamefont\endcsname\relax
  \def\bibfnamefont#1{#1}\fi
\expandafter\ifx\csname citenamefont\endcsname\relax
  \def\citenamefont#1{#1}\fi
\expandafter\ifx\csname url\endcsname\relax
  \def\url#1{\texttt{#1}}\fi
\expandafter\ifx\csname urlprefix\endcsname\relax\def\urlprefix{URL }\fi
\providecommand{\bibinfo}[2]{#2}
\providecommand{\eprint}[2][]{\url{#2}}

\bibitem[{\citenamefont{Friar et~al.}(1990)\citenamefont{Friar, Gibson, Payne,
  Bernstein, and Chupp}}]{fri90}
\bibinfo{author}{\bibfnamefont{J.}~\bibnamefont{Friar}},
  \bibinfo{author}{\bibfnamefont{B.}~\bibnamefont{Gibson}},
  \bibinfo{author}{\bibfnamefont{G.}~\bibnamefont{Payne}},
  \bibinfo{author}{\bibfnamefont{A.}~\bibnamefont{Bernstein}},
  \bibnamefont{and} \bibinfo{author}{\bibfnamefont{T.~E.} \bibnamefont{Chupp}},
  \bibinfo{journal}{Phys. Rev. C} \textbf{\bibinfo{volume}{42}},
  \bibinfo{pages}{2310} (\bibinfo{year}{1990}).

\bibitem[{\citenamefont{Anthony et~al.}(1993)\citenamefont{Anthony, Arnold,
  Band, Borel, Bosted, Breton, Cates, Chupp, Dietrich, Dunne
  et~al.}}]{PhysRevLett.71.959}
\bibinfo{author}{\bibfnamefont{P.~L.} \bibnamefont{Anthony}},
  \bibinfo{author}{\bibfnamefont{R.~G.} \bibnamefont{Arnold}},
  \bibinfo{author}{\bibfnamefont{H.~R.} \bibnamefont{Band}},
  \bibinfo{author}{\bibfnamefont{H.}~\bibnamefont{Borel}},
  \bibinfo{author}{\bibfnamefont{P.~E.} \bibnamefont{Bosted}},
  \bibinfo{author}{\bibfnamefont{V.}~\bibnamefont{Breton}},
  \bibinfo{author}{\bibfnamefont{G.~D.} \bibnamefont{Cates}},
  \bibinfo{author}{\bibfnamefont{T.~E.} \bibnamefont{Chupp}},
  \bibinfo{author}{\bibfnamefont{F.~S.} \bibnamefont{Dietrich}},
  \bibinfo{author}{\bibfnamefont{J.}~\bibnamefont{Dunne}},
  \bibnamefont{et~al.}, \bibinfo{journal}{Phys. Rev. Lett.}
  \textbf{\bibinfo{volume}{71}}, \bibinfo{pages}{959} (\bibinfo{year}{1993}).

\bibitem[{\citenamefont{Riordan et~al.}(2010)\citenamefont{Riordan, Abrahamyan,
  Craver, Kelleher, Kolarkar, Miller, Cates, Liyanage, Wojtsekhowski, Acha
  et~al.}}]{PhysRevLett.105.262302}
\bibinfo{author}{\bibfnamefont{S.}~\bibnamefont{Riordan}},
  \bibinfo{author}{\bibfnamefont{S.}~\bibnamefont{Abrahamyan}},
  \bibinfo{author}{\bibfnamefont{B.}~\bibnamefont{Craver}},
  \bibinfo{author}{\bibfnamefont{A.}~\bibnamefont{Kelleher}},
  \bibinfo{author}{\bibfnamefont{A.}~\bibnamefont{Kolarkar}},
  \bibinfo{author}{\bibfnamefont{J.}~\bibnamefont{Miller}},
  \bibinfo{author}{\bibfnamefont{G.~D.} \bibnamefont{Cates}},
  \bibinfo{author}{\bibfnamefont{N.}~\bibnamefont{Liyanage}},
  \bibinfo{author}{\bibfnamefont{B.}~\bibnamefont{Wojtsekhowski}},
  \bibinfo{author}{\bibfnamefont{A.}~\bibnamefont{Acha}}, \bibnamefont{et~al.},
  \bibinfo{journal}{Phys. Rev. Lett.} \textbf{\bibinfo{volume}{105}},
  \bibinfo{pages}{262302} (\bibinfo{year}{2010}).

\bibitem[{\citenamefont{Coulter et~al.}(1990)\citenamefont{Coulter, Chupp,
  McDonald, Bowman, Bowman, Szymanski, Yuan, Cates, Benton, and
  Earle}}]{Coulter1990463}
\bibinfo{author}{\bibfnamefont{K.~P.} \bibnamefont{Coulter}},
  \bibinfo{author}{\bibfnamefont{T.~E.} \bibnamefont{Chupp}},
  \bibinfo{author}{\bibfnamefont{A.~B.} \bibnamefont{McDonald}},
  \bibinfo{author}{\bibfnamefont{C.~D.} \bibnamefont{Bowman}},
  \bibinfo{author}{\bibfnamefont{J.~D.} \bibnamefont{Bowman}},
  \bibinfo{author}{\bibfnamefont{J.~J.} \bibnamefont{Szymanski}},
  \bibinfo{author}{\bibfnamefont{V.}~\bibnamefont{Yuan}},
  \bibinfo{author}{\bibfnamefont{G.~D.} \bibnamefont{Cates}},
  \bibinfo{author}{\bibfnamefont{D.~R.} \bibnamefont{Benton}},
  \bibnamefont{and} \bibinfo{author}{\bibfnamefont{E.~D.} \bibnamefont{Earle}},
  \bibinfo{journal}{Nuclear Instruments and Methods in Physics Research Section
  A: Accelerators, Spectrometers, Detectors and Associated Equipment}
  \textbf{\bibinfo{volume}{288}}, \bibinfo{pages}{463 } (\bibinfo{year}{1990}).

\bibitem[{\citenamefont{Albert et~al.}(1994)\citenamefont{Albert, Cates,
  Driehuys, Happer, Saam, Springer, and Wishnia}}]{Nature94}
\bibinfo{author}{\bibfnamefont{M.~S.} \bibnamefont{Albert}},
  \bibinfo{author}{\bibfnamefont{G.~D.} \bibnamefont{Cates}},
  \bibinfo{author}{\bibfnamefont{B.}~\bibnamefont{Driehuys}},
  \bibinfo{author}{\bibfnamefont{W.}~\bibnamefont{Happer}},
  \bibinfo{author}{\bibfnamefont{B.}~\bibnamefont{Saam}},
  \bibinfo{author}{\bibfnamefont{C.~S.} \bibnamefont{Springer}},
  \bibnamefont{and} \bibinfo{author}{\bibfnamefont{A.}~\bibnamefont{Wishnia}},
  \bibinfo{journal}{Nature} \textbf{\bibinfo{volume}{370}},
  \bibinfo{pages}{199} (\bibinfo{year}{1994}).

\bibitem[{\citenamefont{Middleton et~al.}(1995)\citenamefont{Middleton, Black,
  Saam, Cates, Cofer, Guenther, Happer, Hedlund, Johnson, Juvan
  et~al.}}]{MRimaging}
\bibinfo{author}{\bibfnamefont{H.}~\bibnamefont{Middleton}},
  \bibinfo{author}{\bibfnamefont{R.~D.} \bibnamefont{Black}},
  \bibinfo{author}{\bibfnamefont{B.~T.} \bibnamefont{Saam}},
  \bibinfo{author}{\bibfnamefont{G.~D.} \bibnamefont{Cates}},
  \bibinfo{author}{\bibfnamefont{G.~P.} \bibnamefont{Cofer}},
  \bibinfo{author}{\bibfnamefont{R.}~\bibnamefont{Guenther}},
  \bibinfo{author}{\bibfnamefont{W.}~\bibnamefont{Happer}},
  \bibinfo{author}{\bibfnamefont{L.~W.} \bibnamefont{Hedlund}},
  \bibinfo{author}{\bibfnamefont{G.~A.} \bibnamefont{Johnson}},
  \bibinfo{author}{\bibfnamefont{K.}~\bibnamefont{Juvan}},
  \bibnamefont{et~al.}, \bibinfo{journal}{Magnetic Resonance in Medicine}
  \textbf{\bibinfo{volume}{33}}, \bibinfo{pages}{271} (\bibinfo{year}{1995}).

\bibitem[{\citenamefont{Colegrove et~al.}(1963)\citenamefont{Colegrove,
  Schearer, and G.Walters}}]{colegrove1963}
\bibinfo{author}{\bibfnamefont{F.}~\bibnamefont{Colegrove}},
  \bibinfo{author}{\bibfnamefont{L.}~\bibnamefont{Schearer}}, \bibnamefont{and}
  \bibinfo{author}{\bibnamefont{G.Walters}}, \bibinfo{journal}{Physical Review}
  \textbf{\bibinfo{volume}{132}}, \bibinfo{pages}{2561} (\bibinfo{year}{1963}).

\bibitem[{\citenamefont{Nacher and M.Leduc}(1985)}]{nacher1985}
\bibinfo{author}{\bibfnamefont{P.}~\bibnamefont{Nacher}} \bibnamefont{and}
  \bibinfo{author}{\bibnamefont{M.Leduc}}, \bibinfo{journal}{Journal de
  Physique} \textbf{\bibinfo{volume}{46}}, \bibinfo{pages}{2057}
  (\bibinfo{year}{1985}).

\bibitem[{\citenamefont{Bouchiat et~al.}(1960)\citenamefont{Bouchiat, Carver,
  and Varnum}}]{bouchiat1960}
\bibinfo{author}{\bibfnamefont{M.}~\bibnamefont{Bouchiat}},
  \bibinfo{author}{\bibfnamefont{T.}~\bibnamefont{Carver}}, \bibnamefont{and}
  \bibinfo{author}{\bibfnamefont{C.}~\bibnamefont{Varnum}},
  \bibinfo{journal}{Physical Review Letters} \textbf{\bibinfo{volume}{5}},
  \bibinfo{pages}{373} (\bibinfo{year}{1960}).

\bibitem[{\citenamefont{Bhaskar et~al.}(1982)\citenamefont{Bhaskar, Happer, and
  McClelland}}]{bhaskar1982}
\bibinfo{author}{\bibfnamefont{N.}~\bibnamefont{Bhaskar}},
  \bibinfo{author}{\bibfnamefont{W.}~\bibnamefont{Happer}}, \bibnamefont{and}
  \bibinfo{author}{\bibfnamefont{T.}~\bibnamefont{McClelland}},
  \bibinfo{journal}{Physical Review Letters} \textbf{\bibinfo{volume}{49}},
  \bibinfo{pages}{25} (\bibinfo{year}{1982}).

\bibitem[{\citenamefont{Chupp et~al.}(1987)\citenamefont{Chupp, Wagshul,
  Coulter, McDonald, and Happer}}]{PhysRevC.36.2244}
\bibinfo{author}{\bibfnamefont{T.~E.} \bibnamefont{Chupp}},
  \bibinfo{author}{\bibfnamefont{M.~E.} \bibnamefont{Wagshul}},
  \bibinfo{author}{\bibfnamefont{K.~P.} \bibnamefont{Coulter}},
  \bibinfo{author}{\bibfnamefont{A.~B.} \bibnamefont{McDonald}},
  \bibnamefont{and} \bibinfo{author}{\bibfnamefont{W.}~\bibnamefont{Happer}},
  \bibinfo{journal}{Phys. Rev. C} \textbf{\bibinfo{volume}{36}},
  \bibinfo{pages}{2244} (\bibinfo{year}{1987}).

\bibitem[{\citenamefont{Krimmer et~al.}(2009)\citenamefont{Krimmer, Distler,
  Heil, Karpuk, Kiselev, Salhi, and Otten}}]{Krimmer200918}
\bibinfo{author}{\bibfnamefont{J.}~\bibnamefont{Krimmer}},
  \bibinfo{author}{\bibfnamefont{M.}~\bibnamefont{Distler}},
  \bibinfo{author}{\bibfnamefont{W.}~\bibnamefont{Heil}},
  \bibinfo{author}{\bibfnamefont{S.}~\bibnamefont{Karpuk}},
  \bibinfo{author}{\bibfnamefont{D.}~\bibnamefont{Kiselev}},
  \bibinfo{author}{\bibfnamefont{Z.}~\bibnamefont{Salhi}}, \bibnamefont{and}
  \bibinfo{author}{\bibfnamefont{E.}~\bibnamefont{Otten}},
  \bibinfo{journal}{Nuclear Instruments and Methods in Physics Research Section
  A: Accelerators, Spectrometers, Detectors and Associated Equipment}
  \textbf{\bibinfo{volume}{611}}, \bibinfo{pages}{18 } (\bibinfo{year}{2009}).

\bibitem[{\citenamefont{Chupp et~al.}(1992)\citenamefont{Chupp, Loveman,
  Thompson, Bernstein, and Tieger}}]{PhysRevC.45.915}
\bibinfo{author}{\bibfnamefont{T.~E.} \bibnamefont{Chupp}},
  \bibinfo{author}{\bibfnamefont{R.~A.} \bibnamefont{Loveman}},
  \bibinfo{author}{\bibfnamefont{A.~K.} \bibnamefont{Thompson}},
  \bibinfo{author}{\bibfnamefont{A.~M.} \bibnamefont{Bernstein}},
  \bibnamefont{and} \bibinfo{author}{\bibfnamefont{D.~R.}
  \bibnamefont{Tieger}}, \bibinfo{journal}{Phys. Rev. C}
  \textbf{\bibinfo{volume}{45}}, \bibinfo{pages}{915} (\bibinfo{year}{1992}).

\bibitem[{\citenamefont{Happer et~al.}(2001)\citenamefont{Happer, Cates,
  Romalis, and Erickson}}]{happer2001}
\bibinfo{author}{\bibfnamefont{W.}~\bibnamefont{Happer}},
  \bibinfo{author}{\bibfnamefont{G.}~\bibnamefont{Cates}},
  \bibinfo{author}{\bibfnamefont{M.}~\bibnamefont{Romalis}}, \bibnamefont{and}
  \bibinfo{author}{\bibfnamefont{C.}~\bibnamefont{Erickson}},
  \emph{\bibinfo{title}{U.\uppercase{S}. \uppercase{P}atent \uppercase{N}o.
  6,318,092}} (\bibinfo{year}{2001}).

\bibitem[{\citenamefont{Babcock et~al.}(2003)\citenamefont{Babcock, Nelson,
  Kadlecek, Driehuys, Anderson, Hersman, and Walker}}]{babcock2003}
\bibinfo{author}{\bibfnamefont{E.}~\bibnamefont{Babcock}},
  \bibinfo{author}{\bibfnamefont{I.}~\bibnamefont{Nelson}},
  \bibinfo{author}{\bibfnamefont{S.}~\bibnamefont{Kadlecek}},
  \bibinfo{author}{\bibfnamefont{B.}~\bibnamefont{Driehuys}},
  \bibinfo{author}{\bibfnamefont{L.~W.} \bibnamefont{Anderson}},
  \bibinfo{author}{\bibfnamefont{F.~W.} \bibnamefont{Hersman}},
  \bibnamefont{and} \bibinfo{author}{\bibfnamefont{T.~G.}
  \bibnamefont{Walker}}, \bibinfo{journal}{Phys. Rev. Lett.}
  \textbf{\bibinfo{volume}{91}}, \bibinfo{pages}{123003}
  (\bibinfo{year}{2003}).

\bibitem[{\citenamefont{Wojtsekhowski}(2002)}]{woj02}
\bibinfo{author}{\bibfnamefont{B.}~\bibnamefont{Wojtsekhowski}}
  (\bibinfo{publisher}{World Scientific}, \bibinfo{year}{2002}),
  \bibinfo{note}{in Proceedings of Exclusive Processes At High Momentum
  Transfer, JLAB-PHY-02-37}.

\bibitem[{\citenamefont{Babcock et~al.}(2006)\citenamefont{Babcock, Chann,
  Walker, Chen, and Gentile}}]{babcock2006}
\bibinfo{author}{\bibfnamefont{E.}~\bibnamefont{Babcock}},
  \bibinfo{author}{\bibfnamefont{B.}~\bibnamefont{Chann}},
  \bibinfo{author}{\bibfnamefont{T.~G.} \bibnamefont{Walker}},
  \bibinfo{author}{\bibfnamefont{W.~C.} \bibnamefont{Chen}}, \bibnamefont{and}
  \bibinfo{author}{\bibfnamefont{T.~R.} \bibnamefont{Gentile}},
  \bibinfo{journal}{Phys. Rev. Lett.} \textbf{\bibinfo{volume}{96}},
  \bibinfo{pages}{083003} (\bibinfo{year}{2006}).

\bibitem[{\citenamefont{Newbury et~al.}(1993)\citenamefont{Newbury, Barton,
  Cates, Happer, and Middleton}}]{new93}
\bibinfo{author}{\bibfnamefont{N.~R.} \bibnamefont{Newbury}},
  \bibinfo{author}{\bibfnamefont{A.~S.} \bibnamefont{Barton}},
  \bibinfo{author}{\bibfnamefont{G.~D.} \bibnamefont{Cates}},
  \bibinfo{author}{\bibfnamefont{W.}~\bibnamefont{Happer}}, \bibnamefont{and}
  \bibinfo{author}{\bibfnamefont{H.}~\bibnamefont{Middleton}},
  \bibinfo{journal}{Phys. Rev. A} \textbf{\bibinfo{volume}{48}},
  \bibinfo{pages}{4411} (\bibinfo{year}{1993}).

\bibitem[{\citenamefont{Bonin et~al.}(1988{\natexlab{a}})\citenamefont{Bonin,
  Walker, and Happer}}]{PhysRevA.37.3270}
\bibinfo{author}{\bibfnamefont{K.~D.} \bibnamefont{Bonin}},
  \bibinfo{author}{\bibfnamefont{T.~G.} \bibnamefont{Walker}},
  \bibnamefont{and} \bibinfo{author}{\bibfnamefont{W.}~\bibnamefont{Happer}},
  \bibinfo{journal}{Phys. Rev. A} \textbf{\bibinfo{volume}{37}},
  \bibinfo{pages}{3270} (\bibinfo{year}{1988}{\natexlab{a}}).

\bibitem[{\citenamefont{Bonin et~al.}(1988{\natexlab{b}})\citenamefont{Bonin,
  Saltzberg, and Happer}}]{PhysRevA.38.4481}
\bibinfo{author}{\bibfnamefont{K.~D.} \bibnamefont{Bonin}},
  \bibinfo{author}{\bibfnamefont{D.~P.} \bibnamefont{Saltzberg}},
  \bibnamefont{and} \bibinfo{author}{\bibfnamefont{W.}~\bibnamefont{Happer}},
  \bibinfo{journal}{Phys. Rev. A} \textbf{\bibinfo{volume}{38}},
  \bibinfo{pages}{4481} (\bibinfo{year}{1988}{\natexlab{b}}).

\bibitem[{\citenamefont{Coulter et~al.}(1989)\citenamefont{Coulter, McDonald,
  Cates, Happer, and Chupp}}]{Coulter198929}
\bibinfo{author}{\bibfnamefont{K.~P.} \bibnamefont{Coulter}},
  \bibinfo{author}{\bibfnamefont{A.~B.} \bibnamefont{McDonald}},
  \bibinfo{author}{\bibfnamefont{G.~D.} \bibnamefont{Cates}},
  \bibinfo{author}{\bibfnamefont{W.}~\bibnamefont{Happer}}, \bibnamefont{and}
  \bibinfo{author}{\bibfnamefont{T.~E.} \bibnamefont{Chupp}},
  \bibinfo{journal}{Nuclear Instruments and Methods in Physics Research Section
  A: Accelerators, Spectrometers, Detectors and Associated Equipment}
  \textbf{\bibinfo{volume}{276}}, \bibinfo{pages}{29 } (\bibinfo{year}{1989}).

\bibitem[{\citenamefont{Jones and \textit{et al, }}(1993)}]{jon93}
\bibinfo{author}{\bibfnamefont{C.}~\bibnamefont{Jones}} \bibnamefont{and}
  \bibinfo{author}{\bibnamefont{\textit{et al, }}}, \bibinfo{journal}{Physical
  Review C} \textbf{\bibinfo{volume}{47}}, \bibinfo{pages}{110}
  (\bibinfo{year}{1993}).

\bibitem[{\citenamefont{Kominis}(2001)}]{kominis}
\bibinfo{author}{\bibfnamefont{I.}~\bibnamefont{Kominis}}, Ph.D. thesis,
  \bibinfo{school}{Princeton University} (\bibinfo{year}{2001}).

\bibitem[{\citenamefont{Abragam}(1961)}]{Abragam}
\bibinfo{author}{\bibfnamefont{A.}~\bibnamefont{Abragam}},
  \emph{\bibinfo{title}{Principles of Nuclear Magnetism}}
  (\bibinfo{publisher}{Oxford University Press}, \bibinfo{year}{1961}).

\bibitem[{\citenamefont{Romalis}(1997)}]{romulus}
\bibinfo{author}{\bibfnamefont{M.}~\bibnamefont{Romalis}}, Ph.D. thesis,
  \bibinfo{school}{Princeton University} (\bibinfo{year}{1997}).

\bibitem[{\citenamefont{Zheng}(2002)}]{Zheng}
\bibinfo{author}{\bibfnamefont{X.}~\bibnamefont{Zheng}}, Ph.D. thesis,
  \bibinfo{school}{M.I.T.} (\bibinfo{year}{2002}).

\bibitem[{\citenamefont{Kestin et~al.}(1984{\natexlab{a}})\citenamefont{Kestin,
  Knierim, Mason, Najafi, Ro, and Waldman}}]{JPCRD}
\bibinfo{author}{\bibfnamefont{J.}~\bibnamefont{Kestin}},
  \bibinfo{author}{\bibfnamefont{K.}~\bibnamefont{Knierim}},
  \bibinfo{author}{\bibfnamefont{E.~A.} \bibnamefont{Mason}},
  \bibinfo{author}{\bibfnamefont{B.}~\bibnamefont{Najafi}},
  \bibinfo{author}{\bibfnamefont{S.~T.} \bibnamefont{Ro}}, \bibnamefont{and}
  \bibinfo{author}{\bibfnamefont{M.}~\bibnamefont{Waldman}},
  \bibinfo{journal}{J. Phys. Chem. Ref. Data} \textbf{\bibinfo{volume}{13}}
  (\bibinfo{year}{1984}{\natexlab{a}}).

\bibitem[{\citenamefont{Garwin and Reich}(1959)}]{PhysRev.115.1478}
\bibinfo{author}{\bibfnamefont{R.~L.} \bibnamefont{Garwin}} \bibnamefont{and}
  \bibinfo{author}{\bibfnamefont{H.~A.} \bibnamefont{Reich}},
  \bibinfo{journal}{Phys. Rev.} \textbf{\bibinfo{volume}{115}},
  \bibinfo{pages}{1478} (\bibinfo{year}{1959}).

\bibitem[{\citenamefont{Fitzsimmons et~al.}(1969)\citenamefont{Fitzsimmons,
  Tankersley, and Walters}}]{fit69}
\bibinfo{author}{\bibfnamefont{W.}~\bibnamefont{Fitzsimmons}},
  \bibinfo{author}{\bibfnamefont{L.}~\bibnamefont{Tankersley}},
  \bibnamefont{and} \bibinfo{author}{\bibfnamefont{G.}~\bibnamefont{Walters}},
  \bibinfo{journal}{Physical Review} \textbf{\bibinfo{volume}{179}},
  \bibinfo{pages}{156} (\bibinfo{year}{1969}).

\bibitem[{\citenamefont{Anger et~al.}(2008)\citenamefont{Anger, Schrank,
  Schoeck, Butler, Solum, Pugmire, and Saam}}]{ang08}
\bibinfo{author}{\bibfnamefont{B.~C.} \bibnamefont{Anger}},
  \bibinfo{author}{\bibfnamefont{G.}~\bibnamefont{Schrank}},
  \bibinfo{author}{\bibfnamefont{A.}~\bibnamefont{Schoeck}},
  \bibinfo{author}{\bibfnamefont{K.~A.} \bibnamefont{Butler}},
  \bibinfo{author}{\bibfnamefont{M.~S.} \bibnamefont{Solum}},
  \bibinfo{author}{\bibfnamefont{R.~J.} \bibnamefont{Pugmire}},
  \bibnamefont{and} \bibinfo{author}{\bibfnamefont{B.}~\bibnamefont{Saam}},
  \bibinfo{journal}{Physical Review A} \textbf{\bibinfo{volume}{78}},
  \bibinfo{pages}{043406} (\bibinfo{year}{2008}).

\bibitem[{\citenamefont{Romalis and Cates}(1998)}]{romalis1998}
\bibinfo{author}{\bibfnamefont{M.~V.} \bibnamefont{Romalis}} \bibnamefont{and}
  \bibinfo{author}{\bibfnamefont{G.~D.} \bibnamefont{Cates}},
  \bibinfo{journal}{Phys. Rev. A.} \textbf{\bibinfo{volume}{105}},
  \bibinfo{pages}{3004} (\bibinfo{year}{1998}).

\bibitem[{\citenamefont{Cates et~al.}(2009)\citenamefont{Cates, Riordan, and
  Wojtsekhowski}}]{GEN2prop}
\bibinfo{author}{\bibfnamefont{G.}~\bibnamefont{Cates}},
  \bibinfo{author}{\bibfnamefont{S.}~\bibnamefont{Riordan}}, \bibnamefont{and}
  \bibinfo{author}{\bibfnamefont{B.}~\bibnamefont{Wojtsekhowski}}
  (\bibinfo{year}{2009}), \bibinfo{note}{\uppercase{J}Lab experiment
  E12-06-016}.

\bibitem[{\citenamefont{Dodge and Thompson}(1937)}]{dodgeNthompson}
\bibinfo{author}{\bibfnamefont{R.}~\bibnamefont{Dodge}} \bibnamefont{and}
  \bibinfo{author}{\bibfnamefont{M.}~\bibnamefont{Thompson}},
  \emph{\bibinfo{title}{Fluid Mechanics}} (\bibinfo{publisher}{McGraw-Hill Book
  Company, INC.}, \bibinfo{year}{1937}).

\bibitem[{\citenamefont{Tritton}(1977)}]{tritton}
\bibinfo{author}{\bibfnamefont{D.}~\bibnamefont{Tritton}},
  \emph{\bibinfo{title}{Physical Fluid Dynamics}} (\bibinfo{publisher}{Van
  Nostrand Reinhold Company, Ltd.}, \bibinfo{year}{1977}).

\bibitem[{\citenamefont{Qian and \textit{et al.}}(2011)}]{qia11}
\bibinfo{author}{\bibfnamefont{X.}~\bibnamefont{Qian}} \bibnamefont{and}
  \bibinfo{author}{\bibnamefont{\textit{et al.}}}, \bibinfo{journal}{Phys. Rev.
  Lett.} \textbf{\bibinfo{volume}{107}}, \bibinfo{pages}{072003}
  (\bibinfo{year}{2011}).

\bibitem[{\citenamefont{Keller}(1957)}]{PhysRev.105.41}
\bibinfo{author}{\bibfnamefont{W.~E.} \bibnamefont{Keller}},
  \bibinfo{journal}{Phys. Rev.} \textbf{\bibinfo{volume}{105}},
  \bibinfo{pages}{41} (\bibinfo{year}{1957}).

\bibitem[{\citenamefont{Kestin et~al.}(1984{\natexlab{b}})\citenamefont{Kestin,
  Knierim, Mason, Najafi, Ro, and Waldman}}]{Kestin}
\bibinfo{author}{\bibfnamefont{J.}~\bibnamefont{Kestin}},
  \bibinfo{author}{\bibfnamefont{K.}~\bibnamefont{Knierim}},
  \bibinfo{author}{\bibfnamefont{E.}~\bibnamefont{Mason}},
  \bibinfo{author}{\bibfnamefont{B.}~\bibnamefont{Najafi}},
  \bibinfo{author}{\bibfnamefont{S.}~\bibnamefont{Ro}}, \bibnamefont{and}
  \bibinfo{author}{\bibfnamefont{M.}~\bibnamefont{Waldman}},
  \bibinfo{journal}{J. Phys. Chem. Ref. Data} \textbf{\bibinfo{volume}{13}},
  \bibinfo{pages}{229} (\bibinfo{year}{1984}{\natexlab{b}}).

\bibitem[{\citenamefont{Schiller}(1922)}]{schiller}
\bibinfo{author}{\bibfnamefont{L.}~\bibnamefont{Schiller}},
  \bibinfo{journal}{Forschungsarbeiten Ver. deut. Ing.}
  \textbf{\bibinfo{volume}{248}} (\bibinfo{year}{1922}).

\bibitem[{cra(2009)}]{crane}
\emph{\bibinfo{title}{Flow of fluids through valves, fittings, and pipe}},
  Technical Paper No. 410 (\bibinfo{publisher}{Crane Company},
  \bibinfo{year}{2009}).

\bibitem[{\citenamefont{Darby}(2001)}]{Darby}
\bibinfo{author}{\bibfnamefont{R.}~\bibnamefont{Darby}},
  \emph{\bibinfo{title}{Chemical Engineering Fluid Mechanics}}
  (\bibinfo{publisher}{Marcel Dekker, Inc.}, \bibinfo{year}{2001}).

\end{thebibliography}


\end{document}